\documentclass[12pt]{article}
\usepackage{amsmath,amssymb,latexsym,epsfig,euscript,multirow,psfrag}
\RequirePackage[sort&compress,square,comma,numbers]{natbib}

\allowdisplaybreaks

\def\ff{f\hspace{-0.3cm}f}

\begin{document}

\begin{titlepage}

\begin{flushright}
FERMILAB-PUB-08-403-T\\
MZ-TH/08-27\\
ZU-TH 15/08\\[0.2cm]
April 14, 2009
\end{flushright}

\vspace{0.2cm}
\begin{center}
\Large\bf
Renormalization-Group Improved Prediction for\\ 
Higgs Production at Hadron Colliders
\end{center}

\vspace{0.2cm}
\begin{center}
{\sc Valentin Ahrens$^a$, Thomas Becher$^b$, Matthias Neubert$^a$ and Li Lin Yang$^a$}\\
\vspace{0.4cm}
{\sl $^a$\,Institut f\"ur Physik (THEP), 
Johannes Gutenberg-Universit\"at\\ 
D--55099 Mainz, Germany\\[0.3cm]
$^b$\,Fermi National Accelerator Laboratory\\
P.O. Box 500, Batavia, IL 60510, U.S.A.}
\end{center}

\vspace{0.2cm}
\begin{abstract}
\vspace{0.2cm}
\noindent 
We use renormalization-group methods in effective field theory to improve the theoretical prediction for the cross section for Higgs-boson production at hadron colliders. In addition to soft-gluon resummation at N$^3$LL, we also resum enhanced contributions of the form $(C_A\pi\alpha_s)^n$, which arise in the analytic continuation of the gluon form factor to time-like momentum transfer. This resummation is achieved by evaluating the matching corrections arising at the Higgs-boson mass scale at a time-like renormalization point $\mu^2<0$, followed by renormalization-group evolution to $\mu^2>0$. We match our resummed result to NNLO fixed-order perturbation theory and give numerical predictions for the total production cross section as a function of the Higgs-boson mass. Resummation effects are significant even at NNLO, where our improved predictions for the cross sections at the Tevatron and the LHC exceed the fixed-order predictions by about 13\% and 8\%, respectively, for $m_H=120$\,GeV. We also discuss the application of our technique to other time-like processes such as Drell-Yan production, $e^+ e^-\to\mbox{hadrons}$, and hadronic decays of the Higgs boson.
\end{abstract}
\vfil

\end{titlepage}

\section{Introduction}

The search for the Higgs boson is of highest priority in the experimental programs at the Tevatron and the LHC. A large effort is thus made to obtain precise theoretical predictions for the corresponding production cross sections. At hadron colliders the dominant production channel is the gluon fusion process through a top-quark loop. The total cross section has been calculated in the heavy top-quark limit up to next-to-next-to-leading order (NNLO) in QCD \cite{Georgi:1977gs,Dawson:1990zj,Djouadi:1991tka,Harlander:2002wh,Anastasiou:2002yz,Ravindran:2003um}, and fully differential predictions are available at the same order for the decays of the Higgs boson into two photons \cite{Anastasiou:2005qj} and four leptons \cite{Anastasiou:2007mz,Grazzini:2008tf}. The exact dependence of the total cross section on the top-quark and Higgs-boson masses is known at NLO \cite{Graudenz:1992pv,Spira:1995rr}.

The perturbative corrections to the total cross section turn out to be
surprisingly large: for a light Higgs boson, the NNLO $K$-factor is around 2, and a scale uncertainty of $\pm (10\mbox{--}15)\%$ remains even at this order. The theoretical prediction has been refined using soft-gluon resummation, which has been implemented at NNLL order \cite{Catani:2003zt} and recently even at the N$^3$LL level \cite{Moch:2005ky,Laenen:2005uz,Idilbi:2005ni,Ravindran:2006cg,Idilbi:2006dg}. The soft-gluon resummation reduces the scale dependence, but the large $K$-factor remains almost unchanged. Indeed, it is not obvious why the cross section should be dominated by soft-gluon radiation: given the large center-of-mass energy of the LHC, there is plenty of phase space available for hard radiation. 

In a recent paper \cite{Ahrens:2008qu}, we have shown that the large $K$-factor is mostly due to terms of the form $(C_A\pi\alpha_s)^n$ in the perturbative series, which arise in the analytic continuation of (double) logarithmic terms in the gluon form factor from space-like to time-like kinematics, $\ln Q^2\to\ln q^2-i\pi$. Being related to Sudakov logarithms, these ``$\pi^2$-enhanced'' contributions can be resummed \cite{Parisi:1979xd,Magnea:1990zb,Bakulev:2000uh,Eynck:2003fn}. Effective field-theory methods provide a particularly simple framework for performing this resummation by implementing matching calculations at time-like momentum transfer and extending renormalization-group (RG) evolution into the complex momentum plane \cite{Ahrens:2008qu}. At first sight it might appear unsystematic to resum $\pi^2$-enhanced perturbative corrections, which cannot be separated from other numerical coefficients (including $\pi^2$ terms not associated with analytic continuation) in a parametric way. However, in our RG framework this resummation simply corresponds to the proper choice of a particular matching scale and as such is unambiguous and physically motivated. The final result for the RG-improved cross section follows straightforwardly by applying the usual rules of effective field theory at every matching step. Our approach to the resummation of $\pi^2$-enhanced corrections is similar in spirit to the analysis of the $e^+ e^-\to\mbox{hadrons}$ cross section using a running coupling evaluated at time-like momentum transfer discussed a long time ago in \cite{Radyushkin:1982kg}, and to the ``contour-improved perturbation theory'' introduced in the analysis of hadronic $\tau$ decays in \cite{Pivovarov:1991rh,Le Diberder:1992te}. It can be viewed as an extension of these methods to problems with Sudakov double logarithms.

In the present paper, we resum both threshold logarithms from soft-gluon emission and the $\pi^2$-enhanced terms using the momentum-space formalism developed in \cite{Becher:2006nr,Becher:2006mr,Becher:2007ty}. Our result is based on the factorization of the cross section near threshold into a hard and a soft function. The resummation is achieved by solving the RG equations for the different parts. In contrast to the standard treatment in Mellin space, this approach yields simple analytic results for the resummed hard-scattering kernels in momentum space and is free of Landau-pole ambiguities. With a phenomenological analysis we investigate to what extent the partonic threshold is enhanced due to the fall-off of the parton distribution functions (PDFs). We find that at LHC energies and for the relevant values of the Higgs-boson mass the scale of the soft emission is not much lower than $m_H$, so that no numerically large logarithms arise from soft emissions. The main numerical effect of RG improvement is thus due to the resummation of the $(C_A\pi\alpha_s)^n$ terms in the virtual corrections. In our RG framework, this resummation is accomplished by evaluating the hard matching corrections at a scale $\mu_h^2=-m_H^2-i\epsilon$ instead of the conventional choice $\mu_h^2=+m_H^2$.

We begin our analysis with a brief review of the fixed-order results for the total cross section and study to which extent the cross section is dominated by the leading singular terms near the partonic threshold. We then discuss the factorization properties of the hard-scattering kernels in the threshold region and derive the formulas for the RG resummation of large perturbative corrections in momentum space. After determining the default values of the matching scales, we present a detailed phenomenological analysis and make predictions for the Higgs-boson production cross sections at the Tevatron and the LHC. Compared with previous studies, we find significantly faster convergence and improved stability of the perturbative expansion. We finally comment on applications of RG-improved perturbation theory to other time-like processes, such as Drell-Yan production, the $e^- e^-\to\mbox{hadrons}$ cross section, and the total hadronic Higgs-boson decay rate. In particular, we explain why the latter two processes do not contain $\pi^2$-enhanced corrections of the type present in Drell-Yan or Higgs-boson production.

\section{Fixed-order results} 
\label{sec:fix}

We consider the production of a Higgs boson in hadron-hadron collisions at center-of-mass energy $\sqrt{s}$. The total cross section can be written as
\begin{equation}\label{sigma0}
   \sigma = \sigma_0 \sum_{i,j} \int_\tau^1\!\frac{dz}{z}\,
   C_{ij}(z,m_t,m_H,\mu_f)\,\ff_{ij}(\tau/z,\mu_f) \,,
\end{equation}
where $\tau=m_H^2/s$, 
\begin{equation}\label{ffdef}
   \ff_{ij}(y,\mu) = \int_y^1\!\frac{dx}{x}\,
   f_{i/N_1}(x,\mu)\,f_{j/N_2}(y/x,\mu)
\end{equation}
are the effective parton luminosities, and $C_{ij}$ are hard-scattering kernels, which are known to NNLO in perturbation theory \cite{Harlander:2002wh,Anastasiou:2002yz,Ravindran:2003um}. The quantity
\begin{equation}\label{sigmaZero}
   \sigma_0 = \frac{G_F}{\sqrt2}\,
    \frac{m_H^2\,\alpha_s^2(\mu_f^2)}{288\pi s}\,
    \Big| \sum_q\,A(x_q) \Big|^2 \,; \qquad
   A(x_q) = \frac{3x_q}{2}\,\big[ 1 + (1-x_q)\,f(x_q) \big]
\end{equation}
with $x_q\equiv 4m_q^2/m_H^2$ 
and 
\begin{equation}
   f(x_q) = \begin{cases} \displaystyle
    \qquad \arcsin^2\frac{1}{\sqrt{x_q}} \,; \quad x_q\ge 1 
    \\[0.4cm] \displaystyle
    - \frac14 \left[ \ln\frac{1+\sqrt{1-x_q}}{1-\sqrt{1-x_q}}
    - i\pi  \right]^2 \,; \quad x_q<1
    \end{cases}
\end{equation}
denotes the Born-level cross section in units of the gluon-gluon luminosity $\ff_{gg}(\tau,\mu_f)$. The function $A(x_q)$ results from a quark loop connecting two gluons with the Higgs boson. It approaches 1 for $x_q\to\infty$ and vanishes proportional to $x_q$ for $x_q\to 0$. It follows that Higgs-boson production is predominantly mediated by a top-quark loop, while the contributions from lighter fermions are strongly suppressed. We include radiative corrections in the heavy top-quark limit, i.e., we will only keep logarithmic top-mass dependence in the hard-scattering kernels $C_{ij}(z,m_t,m_H,\mu_f)$. For not too heavy Higgs-boson masses the terms suppressed by powers of the top-quark mass are numerically very small. Leaving them out greatly simplifies the calculation, since one can then use an effective Lagrangian  obtained after integrating out the top quark.  

Because they are suppressed by $x_q$, the only numerically relevant correction due to lighter fermions is the bottom-quark loop contribution. Its main effect is due to its interference with the top-quark loop, which is well approximated by writing $|\sum_q A(x_q)|^2\approx (1-\epsilon_b)\,|A(x_t)|^2$ with
\begin{equation}\label{epsb}
   \epsilon_b = \frac{3x_b}{4} 
   \left( \ln^2\frac{4}{x_b} - \pi^2 - 4 \right) .
\end{equation}
While using the pole mass is appropriate for the top quark, the virtual $b$-quarks in Higgs-boson production are far off their mass shell, and one should thus use the $\overline{\rm MS}$ quark mass at the Higgs mass scale when evaluating the bottom-quark contribution. At $m_H=120$\,GeV we take $\overline{m}_b\approx 2.8$\,GeV, and the presence of the bottom-quark loop term in (\ref{sigmaZero}) reduces the cross section by $6.5\%$. For comparison, the above approximate treatment would yield $\epsilon_b=7.0\%$. Since the $b$-quark contribution scales like $1/m_H^{2}$, it becomes smaller for higher Higgs-boson masses.

To validate these approximations numerically we have used the computer code \cite{Spira:1995mt}, which includes the exact quark-mass dependence at NLO \cite{Graudenz:1992pv,Spira:1995rr}. For the range $120\,{\rm GeV}<m_H<300\,{\rm GeV}$ we find that the full NLO fixed-order result is about 1\% lower than what is obtained with the above Born-level treatment of finite top-mass effects. The difference is negligible compared to other uncertainties. Also, using the same code we find that the inclusion of the $b$-quark loop decreases the NLO cross section by about $6\%$ at $m_H=120\,{\rm GeV}$ and about $2\%$ at $m_H=300\,{\rm GeV}$, in good agreement with the approximate lowest-order treatment described above.

The variable $z=m_H^2/\hat s$ in (\ref{sigma0}) measures the ratio of the Higgs-boson mass to the parton-parton center-of-mass energy $\sqrt{\hat s}$. The limit $z\to 1$ is referred to as the ``partonic threshold region''. This is the region near Born kinematics, in which the colliding partons have just enough energy to produce the Higgs boson. It is an empirical fact that in many cases this region gives the dominant contributions to the cross section. In Section~\ref{sec:res} we will resum these contributions to all orders in perturbation theory.

The sum in (\ref{sigma0}) extends over all possible combinations of initial partons, but only $C_{gg}$ contributes at the Born level and contains the leading singular terms in the limit $z\to 1$. We split off these terms by writing
\begin{equation}\label{Csplit}
   C_{gg}(z,m_t,m_H,\mu_f) 
   = C(z,m_t,m_H,\mu_f) + C_{gg}^{\rm reg}(z,m_t,m_H,\mu_f) \,,
\end{equation}
where the second piece does not contain singular distributions for $z\to 1$. The explicit expression for the leading singular terms through ${\cal O}(\alpha_s^2)$ is \cite{Dawson:1990zj,Djouadi:1991tka,Harlander:2002wh,Anastasiou:2002yz,Ravindran:2003um}
\begin{equation}\label{fixedorder}
\begin{split}
   C(z,m_t,m_H,\mu_f) 
   &= \delta(1-z) + \frac{\alpha_s}{\pi} \left[ 
    \delta(1-z) \left( \frac{11}{2} + 2\pi^2 \right) 
    + 6 D_1(z) \right] \\
   &\quad\mbox{}+ \left( \frac{\alpha_s}{\pi} \right)^2 
    \left\{ \delta(1-z) \left[ 
    \left( -\frac{13}{3} - \frac{23\pi^2}{6} 
    + \frac{63}{2} \zeta_3 \right) \ln\frac{m_H^2}{\mu_f^2} 
    - \frac{137}{24} \ln\frac{m_t^2}{\mu_f^2} \right.\right. \\
  &\quad\mbox{} \left.\left. + \frac{303}{8} + \frac{349\pi^2}{18} 
   + \frac{23\pi^4}{16} - \frac{307}{12} \zeta_3 \right] 
   - \left( \frac{233}{9} - \frac{23\pi^2}{6} 
   - \frac{351}{2}\,\zeta_3 \right) D_0(z)  \right. \\
   &\qquad\mbox{}+ \left. \bigg( \frac{349}{6} - \frac{15\pi^2}{2}
    - 9\ln^2\frac{m_H^2}{\mu_f^2} \bigg) D_1(z) 
    - \frac{23}{4}\,D_2(z)+ 9 D_3(z) \right\} ,
\end{split}
\end{equation}
where we have defined the distributions
\begin{equation}
  D_n(z) = \bigg[ \frac{1}{1-z}\,
  \ln^n\frac{m_H^2 (1-z)^2}{\mu_f^2 z} \bigg]_+ \,.
\end{equation}
The reason for including a factor $1/z$ in the argument of the logarithm was explained in \cite{Becher:2007ty}. The remaining contributions to the hard-scattering kernels are free of singular distributions. At NLO they read
\begin{equation}\label{Cijreg}
\begin{aligned}
   C_{gg}^{\rm reg}(z,m_t,m_H,\mu_f) 
   &= \frac{\alpha_s}{\pi} \left[ 6 \left( \frac{1}{z} - 2 + z 
    - z^2 \right) \ln\frac{m_H^2 (1-z)^2}{\mu_f^2 z} 
    - \frac{11}{2}\,\frac{(1-z)^3}{z} \right] , \\
   C_{gq}(z,m_t,m_H,\mu_f)  &= \frac{\alpha_s}{\pi} \left[ 
    \frac23\,\frac{1+(1-z)^2}{z} \ln\frac{m_H^2 (1-z)^2}{\mu_f^2 z} 
    + \frac{2z}{3} - \frac{(1-z)^2}{z} \right] , \\
   C_{q\bar q}(z,m_t,m_H,\mu_f)  &= \frac{\alpha_s}{\pi} \left[ 
    \frac{32}{27}\,\frac{(1-z)^3}{z} \right] .
\end{aligned}
\end{equation}
In these expressions $\alpha_s\equiv\alpha_s(\mu_f^2)$, and we do not distinguish between the factorization and renormalization scales. Note that the coefficients of the $\ln[m_H^2(1-z)^2/\mu_f^2 z]$ terms are proportional to the Altarelli-Parisi splitting functions. 

\begin{figure}
\centering
\includegraphics[width=0.49\textwidth]{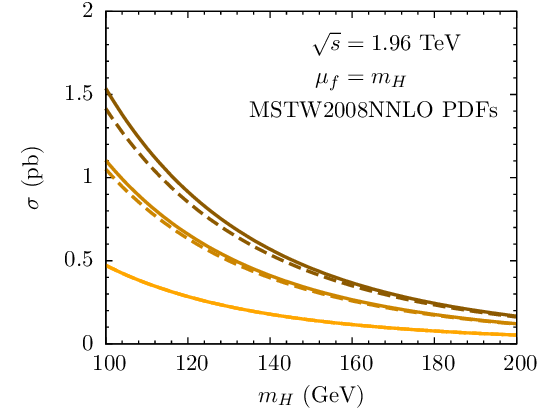}
\includegraphics[width=0.49\textwidth]{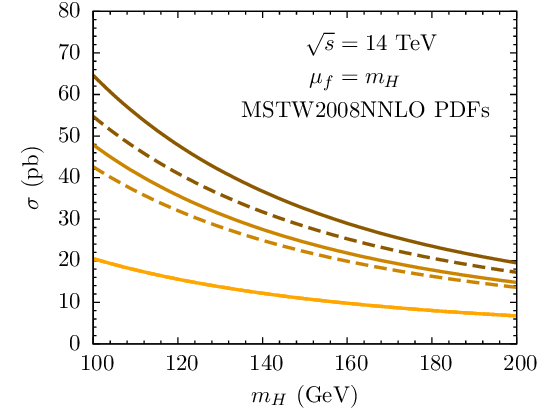}
\caption{\label{fig:leading}
Comparison of the complete fixed-order results (solid lines) and the contributions from the leading singular terms (dashed lines) to the total cross sections for Higgs-boson production at the Tevatron (left) and the LHC (right). We set $\mu_f=m_H$. Darker lines represent higher orders in $\alpha_s$.}
\end{figure}

To visualize the numerical importance of the leading singular terms for Higgs-boson production, we compare the contributions from these terms with the complete fixed-order results in Figure~\ref{fig:leading}. Throughout our analysis we use MSTW2008NNLO PDFs \cite{Martin:2009iq}  and the associated normalization $\alpha_s(m_Z^2)=0.1171\pm 0.0036$  of the running coupling constant, unless noted otherwise. We use three-loop running and $n_f=5$ light quark flavors. The figure shows that the complete fixed-order results are well approximated by the leading singular terms. Taking $m_H=\mu_f=120$\,GeV as an example, the leading singular terms contribute 96\% (94\%) of the NLO (NNLO) cross section at the Tevatron, and 90\% (86\%) of the NLO (NNLO) cross section at the LHC. More specifically, for the LHC this means that 82\% (74\%) of the NLO (NNLO) correction term are captured by the coefficient $C$ in (\ref{fixedorder}). Note also that only $-1\%$ ($-8\%$) of the NLO (NNLO) correction to the cross section are due to parton production channels different from $gg\to H$.

In \cite{Becher:2007ty} we have investigated for the case of Drell-Yan production the question to what extent the dominance of the leading singular terms can be justified based on the strong fall-off of the parton luminosities. In the present case, setting $\mu_f=120$\,GeV for example, we find that $\ff_{gg}(y,\mu_f)\propto y^{-a}$ with $a\approx 2.5$ for $y<0.05$, and $\ff_{gg}(y,\mu_f)\propto (1-y)^b$ with $b\approx 14.5$ for $y>0.3$. Due to this strong fall-off, the integral in (\ref{sigma0}) is dominated by $z$ values near $\tau$. For $\tau$ values exceeding 0.3, the partonic threshold contributions would be enhanced by logarithms of $b\approx 14.5$. However, even at the Tevatron the center-of-mass energy is so high that $\tau\lesssim 0.02$ for Higgs-boson masses below 300\,GeV. In this region the cross section (\ref{sigma0}) is well approximated by the simple formula \cite{Becher:2007ty}
\begin{equation}\label{Born}
   \sigma\approx\sigma_{\rm Born}
    \int_0^1\!dz\,z^{a-1}\,C(z,m_t,m_H,\mu_f) \,; \qquad
   \sigma_{\rm Born} = \sigma_0\,\ff_{gg}(\tau,\mu_f) \,,
\end{equation}
with $a-1\approx 1.5$. Since the weight function $z^{a-1}$ is not strongly peaked near $z=1$, the threshold dominance cannot be explained parametrically in this case. Indeed, we will see later that threshold resummation alone has a very minor effect on the predictions for the cross section. As a side remark, we note that (\ref{Born}) implies the scaling $\sigma\propto m_H^{-2(a-1)}\approx m_H^{-3}$.

Let us now discuss in more detail the different momentum regions that contribute to the Higgs-boson production cross section. For a not too heavy Higgs boson, the gluon-gluon fusion process $gg\to H$ is well approximated by the effective local interaction \cite{Ellis:1975ap,Shifman:1979eb,Vainshtein:1980ea,Inami:1982xt, Voloshin:1985tc}
\begin{equation}\label{Heff}
   {\cal L}_{\rm eff} = C_t(m_t^2,\mu^2)\,\frac{H}{v}\,
 \frac{\alpha_s(\mu^2)}{12\pi} \,
   G_{\mu\nu,a}\,G_a^{\mu\nu} \,,
\end{equation} 
where $v\approx 246$\,GeV is the Higgs vacuum expectation value, and $\mu$ denotes the scale at which the local two-gluon operator is renormalized. The short-distance coefficient $C_t$ is known up to NNNLO \cite{Schroder:2005hy,Chetyrkin:2005ia}. To NNLO, the result reads \cite{Kramer:1996iq,Chetyrkin:1997iv}
\begin{eqnarray}\label{Ct}
   C_t(m_t^2,\mu^2)
   &=& 1 + \frac{\alpha_s(\mu^2)}{4\pi}\,(5C_A-3C_F) \nonumber\\
   &&\mbox{}+ \left( \frac{\alpha_s(\mu^2)}{4\pi} \right)^2
    \bigg[ \frac{27}{2}\,C_F^2 
    + \left( 11\ln\frac{m_t^2}{\mu^2} - \frac{100}{3} \right) C_F C_A 
    - \left( 7\ln\frac{m_t^2}{\mu^2} - \frac{1063}{36} \right) C_A^2 
    \nonumber\\
   &&\quad\mbox{}- \frac{4}{3}\,C_F T_F - \frac{5}{6}\,C_A T_F 
   - \left( 8\ln\frac{m_t^2}{\mu^2} + 5 \right) C_F T_F n_f 
   - \frac{47}{9}\,C_A T_F n_f \bigg] \,.
\end{eqnarray} 
The production cross section is related to the discontinuity of the product of two such effective vertices. As explained earlier, it is a good approximation to keep the exact dependence on the top-quark mass in the Born-level cross section $\sigma_0$ in (\ref{sigma0}), but to employ the effective local interaction (\ref{Heff}) for the analysis of higher-order perturbative corrections. 

\begin{figure}
\centering
\includegraphics[width=0.7\textwidth]{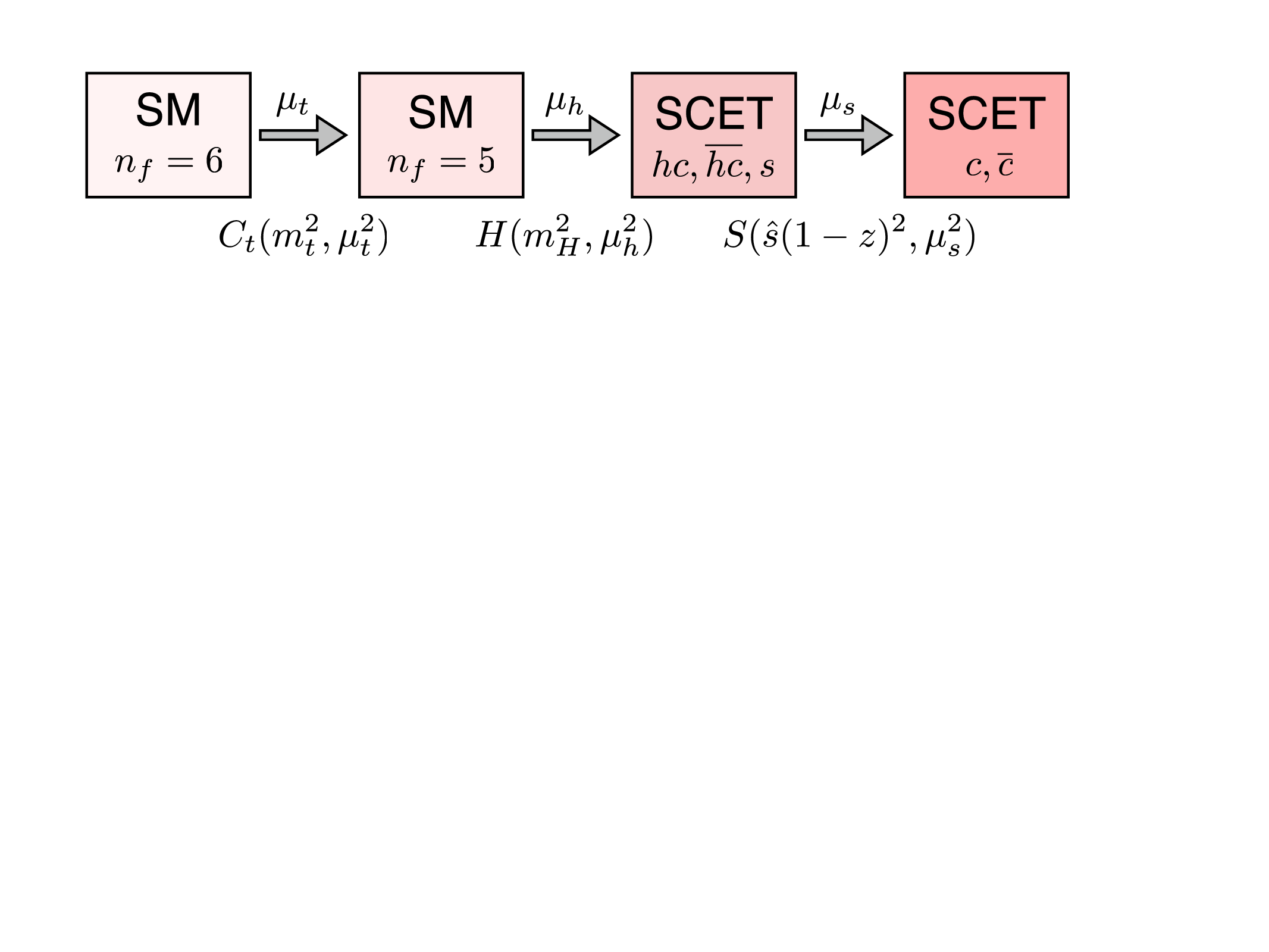}\vspace{-0.3cm}
\caption{\label{fig:matchingsteps} Sequence of matching steps and associated effective theories leading to the factorization theorem (\ref{Cfact}).}
\end{figure}

Once the top quark has been integrated out, the hard-scattering kernels receive contributions associated with two different scales: a ``hard'' scale $\mu_h^2\sim m_H^2$ set by the mass of the Higgs boson, and a ``soft'' scale $\mu_s^2\sim\hat s(1-z)^2=m_H^2(1-z)^2/z$, where $\hat s(1-z)^2/2=(p_\perp^2)_{\rm max}$ is determined by the maximum available transverse momentum. The presence of these two scales is apparent from the structure of the logarithms in the fixed-order terms in (\ref{fixedorder}) and (\ref{Cijreg}), but it can also be derived more rigorously using the method of regions \cite{Beneke:1997zp}. The short-distance coefficient $C$ in (\ref{fixedorder}) can be factorized as
\begin{equation}\label{Cfact}
   C(z,m_t,m_H,\mu_f)
   = \big[C_t(m_t^2,\mu_f^2)\big]^2\,
   H(m_H^2,\mu_f^2)\,S(\hat s(1-z)^2,\mu_f^2) \,,
\end{equation}
and in this way the scale separation becomes explicit. The derivation of this factorization theorem using effective field theory proceeds in analogy with the discussion for the Drell-Yan case presented in \cite{Becher:2007ty} and will not be repeated here in detail. The formula results from a sequence of matching steps illustrated in Figure~\ref{fig:matchingsteps}. The Standard Model with six quark flavors is first matched onto a five-flavor theory by integrating out the heavy top quark. The Wilson coefficient arising in this step is $C_t$. In the next step, the five-flavor Standard Model is matched onto soft-collinear effective theory (SCET) \cite{Bauer:2000yr,Bauer:2001yt,Bauer:2002nz,Beneke:2002ph} containing soft degrees of freedom along with two types of hard-collinear fields aligned with the directions of the particle beams. The corresponding Wilson coefficient is $H$. In the third step, this theory is matched onto another version of SCET, in which the soft modes are integrated out and the hard-collinear modes are replaced by collinear fields of lower virtuality. The soft function $S$ is the matching coefficient arising in this step. The remaining low-energy matrix element is then identified with the parton luminosity function $\ff_{gg}$ defined in (\ref{ffdef}). The calculation of the components $C_t$, $H$, and $S$ at any order in perturbation theory is much simpler than the calculation of the Higgs-boson production cross section at the same order. The factorization formula (\ref{Cfact}) thus provides an approximation to the cross section that requires a minimal amount of calculational work. The all-order resummation of the partonic threshold logarithms (``soft-gluon resummation'') and of other, ``$\pi^2$-enhanced'' terms \cite{Ahrens:2008qu} is then achieved by solving RG equations.

SCET provides field-theoretic definitions of the factors $H$ and $S$ in the factorization formula. The hard function $H$ is the square of the on-shell gluon form factor evaluated at time-like momentum transfer $q^2=m_H^2$, and with infrared divergences subtracted using the $\overline{\rm MS}$ scheme \cite{Idilbi:2006dg,Becher:2006nr,Becher:2007ty}:
\begin{equation}\label{Hdef}
   H(m_H^2,\mu^2) = \left| C_S(-m_H^2-i\epsilon,\mu^2) \right|^2 .
\end{equation}
On a technical level, the function $C_S$ appears as a Wilson coefficient in the matching of the two-gluon operator in (\ref{Heff}) onto an operator in SCET, in which all hard modes have been integrated out. This matching takes the form
\begin{equation}\label{CSdef}
   G_{\mu\nu,a}\,G_a^{\mu\nu}
   \to C_S(Q^2,\mu^2)\,Q^2\,g_{\mu\nu}\,
   {\EuScript A}_{n\perp}^{\mu,a}\,
   {\EuScript A}_{\bar n\perp}^{\nu,a} \,,
\end{equation}
where $Q^2=-q^2$ is (minus) the square of the total momentum carried by the operator. The fields ${\EuScript A}_{n\perp}^{\mu,a}$ and ${\EuScript A}_{\bar n\perp}^{\nu,a}$ are effective, gauge-invariant gluon fields in SCET \cite{Hill:2002vw}. They describe gluons propagating along the two light-like directions $n,\bar n$ defined by the colliding hadrons. The two-loop expression for the Wilson coefficient $C_S$ can be extracted from the results of \cite{Harlander:2000mg}. We write its perturbative series in the form
\begin{equation}\label{CSexp}
   C_S(-m_H^2-i\epsilon,\mu^2) = 1 + \sum_{n=1}^\infty\,c_n(L)
   \left( \frac{\alpha_s(\mu^2)}{4\pi} \right)^n \! ,
\end{equation}
where $L=\ln[(-m_H^2-i\epsilon)/\mu^2]$. The one- and two-loop coefficients read
\begin{equation}\label{c1c2}
\begin{aligned}
   c_1(L) &= C_A \left( -L^2 + \frac{\pi^2}{6} \right) , \\
   c_2(L) &= C_A^2 \left[ \frac{L^4}{2} + \frac{11}{9}\,L^3
    + \left( -\frac{67}{9} + \frac{\pi^2}{6} \right) L^2 
    + \left( \frac{80}{27} - \frac{11\pi^2}{9} - 2\zeta_3 \right) L
    \right. \\
   &\qquad + \left. \frac{5105}{162} + \frac{67\pi^2}{36}
    + \frac{\pi^4}{72} - \frac{143}{9}\,\zeta_3 \right]
    + C_F T_F n_f \left( 4L - \frac{67}{3} + 16\zeta_3 \right) \\
  &\quad\mbox{}+ C_A T_F n_f \left[ -\frac{4}{9}\,L^3
   + \frac{20}{9}\,L^2 
   + \left( \frac{104}{27} + \frac{4\pi^2}{9} \right) L 
   - \frac{1832}{81} - \frac{5\pi^2}{9} - \frac{92}{9}\,\zeta_3 
   \right] .
\end{aligned}
\end{equation}

The soft function $S$ in (\ref{Cfact}) is defined in terms of the Fourier transform of a vacuum expectation value of a Wilson loop in the adjoint representation of $SU(N_c)$. In SCET is arises after the decoupling of soft gluons from the hard-collinear and anti-hard-collinear fields describing the partons originating from the colliding beam particles \cite{Becher:2007ty}. The soft function in the case of Higgs-boson production is closely related to an analogous function entering the Drell-Yan cross section \cite{Idilbi:2006dg,Becher:2007ty}. At two-loop order (but not beyond) the two quantities coincide after a simple replacement of color factors. In the notation of the second reference, we have
\begin{equation}\label{softrel}
\begin{aligned}
   S(\hat s(1-z)^2,\mu_f^2) 
   &= \sqrt{\hat s}\,W_{\rm Higgs}(\hat s(1-z)^2,\mu_f^2) \\
   &= \sqrt{\hat s}\,W_{\rm DY}(\hat s(1-z)^2,\mu_f^2) 
    \Big|_{C_F\to C_A} + {\cal O}(\alpha_s^3) \,.
\end{aligned}
\end{equation}
The explicit form of the result can be derived using formulas compiled in Appendix~B of \cite{Becher:2007ty}.

When one inserts the two-loop expressions for the various component functions into (\ref{Cfact}) and expands the product to ${\cal O}(\alpha_s^2)$, one recovers the expression given in (\ref{fixedorder}). In the following section we will discuss how improved perturbative expressions for the component functions can be obtained by solving RG evolution equations with appropriate boundary conditions. In this way one avoids perturbative logarithms arising when the factorization scale $\mu_f$ is chosen different from the characteristic scales $m_t$, $m_H$, or $\sqrt{\hat s}(1-z)$. Even though these logarithms are not particularly large, their resummation has the effect of improving the stability of the prediction with respect to scale variations. More importantly, however, we will also be able to resum the $\pi^2$-enhanced terms in the perturbative expansion related to the time-like kinematics of the Higgs-boson production process. They have been shown to be responsible for the bulk of the large $K$-factors arising in calculations of the Higgs-production cross sections at the Tevatron and the LHC \cite{Ahrens:2008qu}.

\section{Renormalization-group analysis and resummation}
\label{sec:res}

Our formalism for the resummation of large perturbative corrections in Higgs-boson production is based on effective field theory and follows closely our previous analyses of DIS at large $x$ \cite{Becher:2006nr,Becher:2006mr} and Drell-Yan production \cite{Becher:2007ty}. The two key steps of the approach are deriving a factorization formula such as (\ref{Cfact}) valid near the partonic threshold $z\to 1$, and then using the RG directly in momentum space to resum logarithms arising from ratios of the different scales. We stress that the final, RG-improved formula for the cross section follows unambiguously by applying the rules of effective field theory at each step of the derivation.

The Wilson coefficient $C_t$ appearing when the top quark is integrated out satisfies the RG equation
\begin{equation}\label{eq:rgech}
   \frac{d}{d\ln\mu}\,C_t(m_t^2,\mu^2) 
   = \gamma^t(\alpha_s)\,C_t(m_t^2,\mu^2) \,,
    \qquad \mbox{with} \quad
   \gamma^t(\alpha_s) 
   = \alpha_s^2\,\frac{d}{d\alpha_s}\,
   \frac{\beta(\alpha_s)}{\alpha_s^2} \,.
\end{equation}
The fact that the anomalous dimension is related to the QCD $\beta$-function \cite{Inami:1982xt,Grinstein:1988wz} is not surprising, since the two-gluon operator in (\ref{Heff}) is proportional to the Yang-Mills Lagrangian. The evolution equation can be integrated in closed form and leads to
\begin{equation}\label{Ctsol}
   C_t(m_t^2,\mu_f^2)
   = \frac{\beta\big(\alpha_s(\mu_f^2)\big)/\alpha_s^2(\mu_f^2)}%
          {\beta\big(\alpha_s(\mu_t^2)\big)/\alpha_s^2(\mu_t^2)}\,
   C_t(m_t^2,\mu_t^2) \,,        
\end{equation}
where $\mu_t\sim m_t$ is the matching scale at which the top quark is integrated out. 

The Wilson coefficient $C_S$ arising when hard, virtual quantum corrections to the effective two-gluon vertex (\ref{Heff}) are integrated out obeys an evolution equation reflecting the renormalization properties of the effective two-gluon SCET operator on the right-hand side of the matching relation (\ref{CSdef}). It reads \cite{Becher:2006nr}
\begin{equation}\label{CSevol}
   \frac{d}{d\ln\mu}\,C_S(-m_H^2-i\epsilon,\mu^2)
   = \left[ \Gamma_{\rm cusp}^A(\alpha_s)\,
   \ln\frac{-m_H^2-i\epsilon}{\mu^2} + \gamma^S(\alpha_s) \right] 
   C_S(-m_H^2-i\epsilon,\mu^2) \,,
\end{equation}
where $\Gamma_{\rm cusp}^A$ is the cusp anomalous dimension of Wilson lines with light-like segments in the adjoint representation of $SU(N_c)$. It controls the leading Sudakov double logarithms contained in $C_S$ and is known to three-loop order \cite{Vogt:2004mw}. The single-logarithmic evolution is controlled by the anomalous dimension $\gamma^S$, which can be extracted from the infrared divergences of the on-shell gluon form factor \cite{Becher:2006nr}. Using results from \cite{Moch:2005tm} it can be derived to three-loop order. We collect the relevant expressions for the expansion coefficients of the anomalous dimensions in Appendix~\ref{app1}. The general solution to (\ref{CSevol}) is \cite{Neubert:2004dd}
\begin{equation}\label{CSsol}
   C_S(-m_H^2-i\epsilon,\mu_f^2) 
   \!=\! \exp\!\left[ 2S(\mu_h^2,\mu_f^2) 
   - a_\Gamma(\mu_h^2,\mu_f^2)\,\ln\!\frac{-m_H^2-i\epsilon}{\mu_h^2}
   - a_{\gamma^S}(\mu_h^2,\mu_f^2) \right]\! 
   C_S(-m_H^2-i\epsilon,\mu_h^2) ,
\end{equation}
where $\mu_h$ is the hard matching scale. We have introduced the definitions
\begin{equation}\label{Saints}
\begin{aligned}
   S(\nu^2,\mu^2) 
   &= - \int\limits_{\alpha_s(\nu^2)}^{\alpha_s(\mu^2)}\!
    d\alpha\,\frac{\Gamma_{\rm cusp}^A(\alpha)}{\beta(\alpha)}
    \int\limits_{\alpha_s(\nu^2)}^\alpha
    \frac{d\alpha'}{\beta(\alpha')} \,, \\
   a_\Gamma(\nu^2,\mu^2) 
   &= - \int\limits_{\alpha_s(\nu^2)}^{\alpha_s(\mu^2)}\!
    d\alpha\,\frac{\Gamma_{\rm cusp}^A(\alpha)}{\beta(\alpha)} \,, 
\end{aligned}
\end{equation}
and similarly for the function $a_{\gamma^S}$. The perturbative expansions of these functions obtained at NNLO in RG-improved perturbation theory can be found in the Appendix of \cite{Becher:2006mr}. 

The naive choice $\mu_h^2\sim m_H^2$ of the hard matching scale gives rise to large $\pi^2$ terms in the matching condition (\ref{CSexp}), which arise since $L^2=\ln^2[(-m_H^2-i\epsilon)/\mu_h^2]\sim -\pi^2$ and render the perturbative expansion of the hard function $H$ in (\ref{Hdef}) unstable. We have shown in \cite{Ahrens:2008qu} that these $\pi^2$-enhanced terms are to a large extent responsible for the poor perturbative behavior of fixed-order predictions for the Higgs-boson production cross sections at hadron colliders. We can exploit the fact that the solution (\ref{CSsol}) is formally independent of the hard matching scale to avoid the large $\pi^2$ terms in the matching condition by a proper choice of the matching scale. To this end we set $\mu_h^2\sim -m_H^2-i\epsilon$, so that $\ln[(-m_H^2-i\epsilon)/\mu_h^2]$ remains a small parameter. The $\pi^2$-enhanced terms are then resummed to all orders in perturbation theory and appear in the functions $S$ and $a_\Gamma$ in the exponent in (\ref{CSsol}). With this choice, relation (\ref{CSsol}) involves the running coupling $\alpha_s(\mu^2)$ evaluated at {\em negative\/} argument. The definition $\beta(\alpha_s)=d\alpha_s/d\ln\mu$ of the QCD $\beta$-function implies that
\begin{equation}\label{asdef}
   \int_{\alpha_s(\mu^2)}^{\alpha_s(-\mu^2)}\!
   \frac{d\alpha}{\beta(\alpha)}
   = - \frac{i\pi}{2} \,,
\end{equation}
where here and below $\alpha_s(-\mu^2)$ is to be understood with a $-i\epsilon$ prescription. This relation allows us to define the running coupling at time-like argument in terms of that at space-like momentum transfer. At NLO we obtain
\begin{equation}\label{asNLO}
   \frac{\alpha_s(\mu^2)}{\alpha_s(-\mu^2)}
   = 1 - ia(\mu^2) 
   + \frac{\beta_1}{\beta_0}\,\frac{\alpha_s(\mu^2)}{4\pi}\, 
   \ln\left[ 1 - ia(\mu^2) \right] + {\cal O}(\alpha_s^2) \,,
\end{equation}
where we count $a(\mu^2)\equiv\beta_0\alpha_s(\mu^2)/4$ as an ${\cal O}(1)$ parameter. It is important that (\ref{asdef}) is independent of the path. For example, the evolution of the coupling can be performed on a circle with fixed radius in the complex momentum plane, thereby avoiding the region near the origin, where perturbation theory breaks down. Note that the perturbation-theory coupling $\alpha_s(\mu^2)$ is analytic in the complex $\mu^2$-plane with a cut along the negative real axis. It has an unphysical Landau pole at $\mu^2=\Lambda_{\overline{\rm MS}}^2$, which is of no concern to our discussion since we are interested in very large $|\mu^2|$ values. In practice, we obtain $\alpha_s(\mu_h^2)$ for $\mu_h^2<0$ by simply evaluating the three-loop running coupling at negative values of its argument. 

The soft Wilson loop $W_{\rm Higgs}$ in (\ref{softrel}) obeys an integro-differential evolution equation, which is analogous to that for the soft function in Drell-Yan production discussed in \cite{Becher:2007ty}. The general solution to this equation can be obtained using a Laplace transformation \cite{Becher:2006nr}. It can be written with the help of an associated function $\widetilde s_{\rm Higgs}$, which is given by the Laplace transform of the soft Wilson loop at a matching scale $\mu_s$. The solution is then obtained from
\begin{equation}\label{softol}
   \omega\,W_{\rm Higgs}(\omega^2,\mu_f^2) 
   = \exp\left[ -4S(\mu_s^2,\mu_f^2) 
   + 2a_{\gamma^W}(\mu_s^2,\mu_f^2) \right]
   \widetilde s_{\rm Higgs}(\partial_\eta,\mu_s^2)
   \left( \frac{\omega^2}{\mu_s^2} \right)^\eta 
   \frac{e^{-2\gamma_E\eta}}{\Gamma(2\eta)} \,,
\end{equation}
where $\partial_\eta$ denotes a derivative with respect to an auxiliary parameter $\eta$, which is then set to $\eta=2a_\Gamma(\mu_s^2,\mu_f^2)$. As written above, the solution is valid as long as $\eta>0$. From the RG invariance of the Higgs-boson production cross section one can derive a relation between the anomalous dimension $\gamma^W$ entering in the above solution and the anomalous dimensions of the remaining components in the factorization formula for the cross section \cite{Becher:2007ty}. It reads
\begin{equation}
   \gamma^W = \frac{\beta(\alpha_s)}{\alpha_s}
   + \gamma^t + \gamma^S + 2\gamma^B \,,
\end{equation}
where $2\gamma^B$ is coefficient of the $\delta(1-x)$ term in the Altarelli-Parisi splitting function $P_{g\leftarrow g}(x)$. The three-loop expression for this quantity was obtained in \cite{Vogt:2004mw}, and we collect the corresponding expansion coefficients in Appendix~\ref{app1}. At two-loop order, relation (\ref{softrel}) implies that the associated soft function $\widetilde s_{\rm Higgs}$ is obtained from that in the Drell-Yan case by the replacement $C_F\to C_A$. This gives 
\begin{equation}\label{tils}
  \widetilde s_{\rm Higgs}(L,\mu^2) 
  = 1 + \frac{\alpha_s(\mu^2)}{4\pi}\,C_A 
  \left( 2L^2 + \frac{\pi^2}{3} \right) 
  + \left( \frac{\alpha_s(\mu^2)}{4\pi} \right)^2
  \left( C_A^2\,W_A + C_A T_F n_f\,W_f \right) ,
\end{equation}
with
\begin{equation}
\begin{aligned}
   W_A &= 2L^4 - \frac{22}{9}\,L^3 + \frac{134}{9}\,L^2
    + \left( -\frac{808}{27} + 28\zeta_3 \right) L 
    + \frac{2428}{81} + \frac{67\pi^2}{54} - \frac{5\pi^4}{18} 
    - \frac{22}{9}\,\zeta_3 \,, \\
   W_f &= \frac{8}{9}\,L^3 - \frac{40}{9}\,L^2 + \frac{224}{27}\,L
    - \frac{656}{81} - \frac{10\pi^2}{27} + \frac{8}{9}\,\zeta_3 \,.
\end{aligned}
\end{equation}
The result (\ref{tils}) agrees with a corresponding expressions derived in \cite{Idilbi:2006dg}.

Putting everything together, we arrive at our final formula for the RG-improved expression for the hard-scattering coefficient in (\ref{fixedorder}). It can be written in the form
\begin{equation}\label{eq:cc}
\begin{aligned}
   C(z,m_t,m_H,\mu_f)    
   &= \big[C_t(m_t^2,\mu_t^2)\big]^2\,
    \left| C_S(-m_H^2-i\epsilon,\mu_h^2) \right|^2 
    U(m_H,\mu_t,\mu_h,\mu_s,\mu_f) \\
   &\quad\times \frac{z^{-\eta}}{(1-z)^{1-2\eta}}\, 
    \widetilde s_{\rm Higgs}\bigg( \ln\frac{m_H^2(1-z)^2}{\mu_s^2 z} 
    + \partial_\eta, \mu_s^2 \bigg)\,
    \frac{e^{-2\gamma_E\eta}}{\Gamma(2\eta)} \,,
\end{aligned}
\end{equation}
where 
\begin{equation}\label{Uexact}
\begin{aligned}
   U(m_H,\mu_t,\mu_h,\mu_s,\mu_f)
   &= \frac{\alpha_s^2(\mu_s^2)}{\alpha_s^2(\mu_f^2)} \left[ 
    \frac{\beta\big(\alpha_s(\mu_s^2)\big)/\alpha_s^2(\mu_s^2)}%
         {\beta\big(\alpha_s(\mu_t^2)\big)/\alpha_s^2(\mu_t^2)} 
    \right]^2 
    \left| \left( \frac{-m_H^2-i\epsilon}{\mu_h^2} 
    \right)^{-2a_\Gamma(\mu_h^2,\mu_s^2)} \right| \\
   &\quad\times \left| \exp \left[ 4S(\mu_h^2,\mu_s^2) 
    - 2a_{\gamma^S}(\mu_h^2,\mu_s^2) 
    + 4a_{\gamma^B}(\mu_s^2,\mu_f^2) \right] \right| .
\end{aligned}
\end{equation}
Apart from the factor containing the $\beta$-function, which is related to the evolution of the two-gluon operator in (\ref{Heff}), and the ratio of running couplings, which compensates the scale dependence of the Born-level cross section $\sigma_0$ in (\ref{sigma0}), this result is of the same form as the corresponding expression arising in Drell-Yan production and given in equations~(50) and (51) of \cite{Becher:2007ty}.  Some comments on the effect of the resummation of $\pi^2$-enhanced terms  in the Drell-Yan case will be made in Section~\ref{sec:othercases}.

It is instructive to consider the special limit in which all matching scales are set equal to a common scale $\mu_f\sim m_H$, while $\mu_h^2=-\mu_f^2-i\epsilon$ is still chosen in the time-like region. We then obtain \cite{Ahrens:2008qu}
\begin{equation}\label{beauty}
\begin{split}
   \ln U(m_H,\mu_f,-i\mu_f,\mu_f,\mu_f)
   &= \frac{\Gamma_0^A}{2\beta_0^2}\,\bigg\{
    \frac{4\pi}{\alpha_s(m_H^2)}\! 
    \left[ 2a\arctan(a) - \ln(1+a^2) \right] \\
   &\quad\mbox{}+ \left( \frac{\Gamma_1^A}{\Gamma_0^A} 
    - \frac{\beta_1}{\beta_0} - \frac{\gamma_0^S\beta_0}{\Gamma_0^A}
    \right) \ln(1+a^2) \\
   &\quad\mbox{}+ \frac{\beta_1}{4\beta_0}
    \left[ 4\arctan^2(a) - \ln^2(1+a^2) \right] 
    + {\cal O}(\alpha_s) \bigg\} \,, 
\end{split}
\end{equation}
where $a\equiv a(m_H^2)$. Note that the result is $\mu_f$-independent at this order. The expression for the evolution function simplifies considerably if we treat $a(m_H^2)\approx 0.2$ as a parameter of order $\alpha_s$. Using the fact that $\gamma_0^S=0$, we then find
\begin{equation}\label{Ureexpanded}
   \ln U(m_H,\mu_f,-i\mu_f,\mu_f,\mu_f)
   = \frac{\pi^2}{2}\,\Gamma_{\rm cusp}^A[\alpha_s(m_H^2)]
   + {\cal O}(\alpha_s^3) \,.
\end{equation}
This result makes explicit that the $\pi^2$-enhanced corrections are terms of the form $(C_A\pi\alpha_s)^n$ in perturbation theory and exponentiate at leading order. Numerically, setting $\mu_f=m_H=120$\,GeV we obtain $\ln U=\{0.558,0.560,0.561\}$ at LO, NLO, and NNLO from the exact expression for the evolution function derived from (\ref{Uexact}), indicating that the leading-order terms give by far the dominant effect after RG improvement. The analytical expressions (\ref{beauty}) and (\ref{Ureexpanded}) provide accurate approximations to the exact results. The first equation gives $\ln U=0.557$, while the second one yields $\ln U=0.565$. The close agreement of these two numbers shows that the running of the coupling between $\mu_f^2$ and $-\mu_f^2$ is a minor effect compared with the evolution driven by the anomalous dimension of the effective two-gluon operator in (\ref{CSdef}).

The RG-improved prediction for the leading singular contributions to the Higgs-boson production cross section is obtained by integrating the above expression for the resummed kernel $C$ with the gluon-gluon luminosity function $\ff_{gg}$, see (\ref{sigma0}). In order to also account for the remaining contributions to the cross section, we add to this result the fixed-order contributions arising from the non-singular terms in the hard-scattering kernels, which at NLO have been compiled in (\ref{Cijreg}). In the momentum-space approach the subtractions required to avoid double counting of the resummed terms are straightforwardly implemented as \cite{Becher:2006mr,Becher:2007ty}
\begin{equation}\label{eq:combine}
  \sigma^{\rm RGI} 
  = \sigma^{\rm resummed} \Big|_{\mu_t,\mu_h,\mu_s,\mu_f} 
  + \left( \sigma^{\text{fixed order}} \Big|_{\mu_f} 
  - \sigma^{\rm resummed} \Big|_{\mu_t=\mu_h=\mu_s=\mu_f} \right) .
\end{equation}
We have used this prescription to calculate the fixed-order expressions for the two terms on the right-hand side of (\ref{Csplit}). Note that only after this matching step the cross section is formally independent of the factorization scale $\mu_f$.

Traditionally, threshold resummation is performed in moment space rather than momentum space. For the case at hand, one considers moments of the cross section in $\tau=m_H^2/s$ at fixed $m_H$:
\begin{equation}\label{momentdef}
   \sigma_N = \int_0^1\!d\tau\,\tau^{N-1}\,\sigma \,.
\end{equation}
After this Mellin transformation, the convolution integrals in  (\ref{sigma0}) and (\ref{ffdef}) reduce to products of moments. For the gluon contribution to the cross section, one has 
\begin{align}
  \sigma_N = \sigma_0\, C_{N+1}(m_t,m_H,\mu_f) \, f^{g/N_1}_{N+1}(\mu_f) \, f^{g/N_2}_{N+1}(\mu_f)  \, .
\end{align}
Having the analytical result (\ref{softol}) for the RG equation of the soft function at hand allows us to work directly in momentum space. However, to compare to the results obtained using traditional methods, it is instructive to transform our result (\ref{eq:cc}) for the hard-scattering coefficient to moment space. This was discussed in detail in \cite{Becher:2006mr} for DIS and \cite{Becher:2007ty} for Drell-Yan production. The discussion for Higgs production is completely analogous to the Drell-Yan case, but for the numerical discussion below, it will be useful to have explicit formulae also for the present case. To obtain the moment-space result we note that \cite{Becher:2006qw}
\begin{equation}
 \int_0^1\!dz\,z^{N-1}\,S(m_H^2(1-z)^2,\mu^2) 
   = \widetilde s_{\rm Higgs}\bigg(\ln\frac{m_H^2}{\bar N^2\mu^2},\mu^2\bigg)
   + {\cal O}\left(\frac{1}{N}\right)  ,
\end{equation}
with $\bar N=e^{\gamma_E} N$. Solving the associated RG equation and combining it with the hard function we obtain
\begin{equation}
\begin{aligned}\label{CN}
C_N(m_t,m_H,\mu_f) &= \big[C_t(m_t^2,\mu_t^2)\big]^2\,
    \left| C_S(-m_H^2-i\epsilon,\mu_h^2) \right|^2  \\
   &\quad\times 
    U(m_H,\mu_t,\mu_h,\mu_s,\mu_f) \, \widetilde s_{\rm Higgs}\bigg( \ln\frac{m_H^2}{\mu_s^2}+ \partial_\eta, \mu_s^2 \bigg)\,
    \bar{N}^{-2\eta} +{\cal O}\left(\frac{1}{N}\right) .
\end{aligned}
\end{equation}
The result has exactly the same structure as (\ref{eq:cc}). Evaluating the derivatives with respect to $\eta$ produces logarithms $ \ln\frac{m_H^2}{\bar{N}^2\mu_s^2}$, which can be eliminated choosing $\mu_s^2=\frac{m_H^2}{\bar{N}^2}$. For fixed $\mu_s$, the result (\ref{CN}) has a very simple $N$-dependence, and we can Mellin-invert it analytically using
\begin{eqnarray}\label{Mellininv}
   \frac{1}{2\pi i} \int_{c-i\infty}^{c+i\infty}\!dN\, 
   z^{-N} \bar N^{-2\eta} 
   &=& (-\ln z)^{-1+2\eta}\,\frac{e^{-2\gamma_E\eta}}{\Gamma(2\eta)}
    \nonumber\\
   &=& \sqrt{z}\,\frac{z^{-\eta}}{(1-z)^{1-2\eta}}\,
    \frac{e^{-2\gamma_E\eta}}{\Gamma(2\eta)}
    \left[ 1 + O\Big( (1-z)^2 \Big) \right] .
\end{eqnarray}
The main difference to (\ref{eq:cc}) is an overall factor of $\sqrt{z}$, which amounts to a first-order power correction in the threshold region $z\to 1$. The numerical impact of this factor  will be discussed below.

The result (\ref{CN}) can be compared to the expressions used in the traditional formulation of resummation. For DIS and Drell-Yan production, it has been shown in \cite{Becher:2006mr,Becher:2007ty} that the two methods give identical results for the threshold-enhanced terms when expanded to any fixed order in $\alpha_s$. In these papers, an exact relation between the radiation functions appearing in the traditional framework and the anomalous dimensions and Wilson coefficients in the effective theory was derived. For the Higgs case, the corresponding relation reads 
\begin{equation}
   e^{2\gamma_E\nabla}\,\Gamma(1+2\nabla)\,\frac{D(\alpha_s)}{2} 
   = \gamma_W(\alpha_s) + \nabla\ln\widetilde s_{\rm Higgs}(0,\mu^2)
   - \frac{e^{2\gamma_E\nabla}\,\Gamma(1+2\nabla)-1}{\nabla}\,
   \Gamma_{\rm cusp}(\alpha_s) \,,
\end{equation}
where $\alpha_s\equiv\alpha_s(\mu)$, and $\nabla=d/d\ln\mu^2=[\beta(\alpha_s)/2]\,\partial/\partial\alpha_s$. Using this result, we reproduce the perturbative expression for the radiation function $D(\alpha_s)$ given in \cite{Moch:2005ky,Laenen:2005uz} up to third order in~$\alpha_s$. 

\section{Choice of the matching and factorization scales}
\label{sec:scales}

The RG-improved cross section in (\ref{eq:combine}) is formally independent of each of the three matching scales $\mu_t$, $\mu_h$, and $\mu_s$, as well as of the factorization scale $\mu_f$ at which the parton densities are evaluated. However, in practice a residual scale dependence remains due to the truncation of perturbation theory. It is a standard procedure to take this residual dependence on the scales as an estimate of yet unknown higher-order effects. In our analysis we will independently vary the various matching scales about their default values, whose determination we will now discuss. In the spirit of effective field theory, the matching scales should be chosen such that the matching conditions (i.e., the Wilson coefficients evaluated at the matching scales) in (\ref{eq:cc}) have well-behaved perturbative expansions. All large corrections are then resummed into the evolution function $U$. 

The characteristic scale of the top-quark loop integrated out in the construction of the effective local interaction (\ref{Heff}) is the top-quark mass, and we thus take $\mu_t=m_t$ as our default choice for the first matching scale. With this choice, the perturbation series for the matching coefficient $C_t$ is well behaved. Setting $n_f=5$ for the number of light quark flavors, we find
\begin{equation}
   C_t(m_t^2,m_t^2) 
   =  1 + 0.875\,\alpha_s(m_t^2)
   + 0.623\,\alpha_s^2(m_t^2) + \dots\,.
\end{equation}

As mentioned earlier, the most naive choice for the hard matching scale, $\mu_h=m_H$, does not lead to a well-behaved expansion for the hard matching coefficient. We find
\begin{equation}\label{CS1}
   C_S(-m_H^2-i\epsilon,m_H^2) = 1 + 2.749\,\alpha_s(m_H^2) 
   + (4.844+2.071i)\,\alpha_s^2(m_H^2) + \dots \,.
\end{equation}
The origin of the large expansion coefficients can be traced back to the fact that the Sudakov (double) logarithms contained in the coefficients $c_n(L)$ in (\ref{CSexp}) give rise to $\pi^2$ terms when we analytically continue $L\to\ln(m_H^2/\mu_h^2)-i\pi$. The same happens for the coefficient $C_V$ in Drell-Yan production \cite{Parisi:1979xd,Magnea:1990zb} and for other time-like processes \cite{Bakulev:2000uh}. A vastly better behavior is obtained when the matching scale is chosen in the time-like region \cite{Ahrens:2008qu}. This gives (all arguments are defined with a $-i\epsilon$ prescription)
\begin{equation}\label{CS2}
   C_S(-m_H^2,-m_H^2) = 1 + 0.393\,\alpha_s(-m_H^2) 
   - 0.152\,\alpha_s^2(-m_H^2) + \dots \,.
\end{equation}
Note that the values of the strong coupling in the space-like and time-like regions are not very different from each other. For instance, setting $m_H=120$\,GeV we find $\alpha_s(-m_H^2)/\alpha_s(m_H^2)=0.951+0.213i$. It follows that the stark difference between (\ref{CS1}) and (\ref{CS2}) is not due to the evolution of the running coupling between space-like and time-like values of its argument, but rather due to the evolution of the effective two-gluon operator (\ref{CSdef}) driven by its anomalous dimension. In our phenomenological analysis we will thus use $\mu_h^2=-m_H^2$ as our default choice. Then the $\pi^2$-enhanced corrections are resummed into the evolution function $U$ in (\ref{eq:cc}). In order to illustrate the significance of this resummation, we will sometimes use the naive choice $\mu_h^2=m_H^2$ for comparison.

\begin{figure}
\centering
\includegraphics[width=0.49\textwidth]{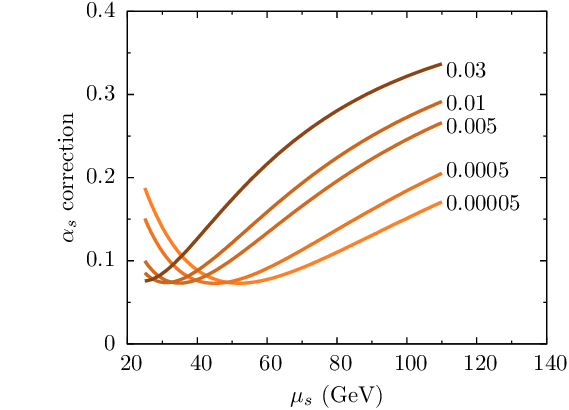}
\includegraphics[width=0.47\textwidth]{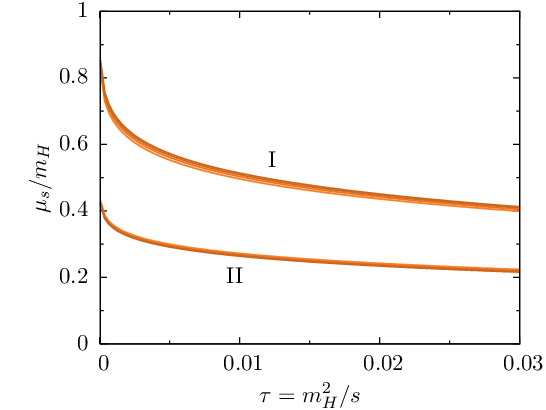}
\caption{\label{fig:soft}
{\em Left:\/} Relative contributions to the total cross section arising from the one-loop corrections to the soft function $\widetilde s_{\rm Higgs}$ as a function of the soft matching scale $\mu_s$, obtained with $\mu_f=m_H=120$\,GeV. The numbers on the curves show the corresponding values of $\tau$. {\em Right:\/} Results for the soft matching scale $\mu_s$ for different values of $\tau$ and four different Higgs-mass values (see text). The upper set of curves corresponds to convergence criterion I, the lower one to criterion II.}
\end{figure}

Let us now discuss the choice of the soft matching scale $\mu_s$, which is non-trivial since the soft function $S$ in (\ref{softrel}) depends on the convolution variable $z$. For the determination of the soft scale we follow the method proposed in \cite{Becher:2007ty}, i.e., we choose the value of $\mu_s$ so that the perturbative expansion of the soft function exhibits a good convergence {\em after\/} the integration over $z$ has been performed. The result thus depends on the process (in particular, on the value of the Higgs-boson mass) and on the shape of the gluon distribution function. The left panel in Figure~\ref{fig:soft} shows the relative contributions to the cross section (normalized to 1) arising from the one-loop terms in the soft function $\widetilde s_{\rm Higgs}$ as a function of $\mu_s$. We choose $\mu_f=m_H=120$~GeV and consider different values of $\tau=m_H^2/s$ between 0.00005 and 0.03, which is the relevant region for our study. The two scale-setting criteria proposed in \cite{Becher:2007ty} are:
\begin{enumerate}
\item[I.] Starting from a high scale, determine the value of $\mu_s$ at which the one-loop correction drops below 15\%.
\item[II.] Choose the value of $\mu_s$ for which the one-loop correction is minimized.
\end{enumerate}
With either choice the two-loop corrections at the corresponding $\mu_s$ values are very small, indicating a good convergence of the perturbative series. The same analysis is repeated for different values of the Higgs-boson mass. The resulting values for the soft scale $\mu_s$ are shown in the right panel in Figure~\ref{fig:soft} for $m_H=120$, 160, 200, and 240\,GeV. Note that the ratio $\mu_s/m_H$ is to a good approximation independent of $m_H$. An analogous scaling behavior was also observed in the Drell-Yan case \cite{Becher:2007ty}. Below we will vary the soft scale between the two choices labeled by $\mu_s^{\rm I}$ and $\mu_s^{\rm II}$ in the figure, taking the average of the two prescriptions as our default value. In practice the values of $\tau$ relevant to Higgs-boson production at the Tevatron or LHC are so small that the ratio $\mu_s/m_H$ can be considered an ${\cal O}(1)$ parameter. This confirms our earlier argument stating that there is no parametric reason to perform soft-gluon (or threshold) resummation in this case.

Following common practice, we take $\mu_f=m_H$ as our default value for the factorization scale. From the perspective of effective field theory it would be more natural to choose $\mu_f$ at or below the soft matching scale $\mu_s$, because the parton densities describe the low-energy hadronic matrix elements after the hard and soft modes have been integrated out. 

\begin{figure}[t]
\centering
\includegraphics[width=0.45\textwidth]{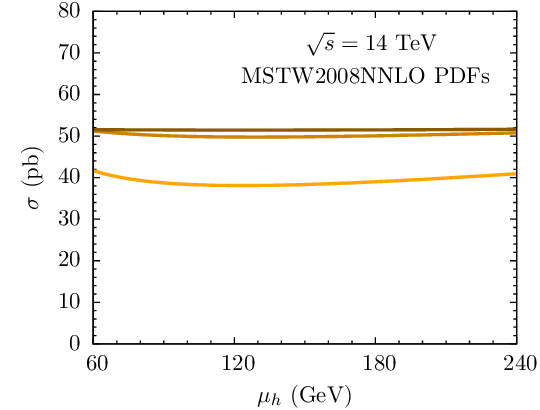}
\includegraphics[width=0.45\textwidth]{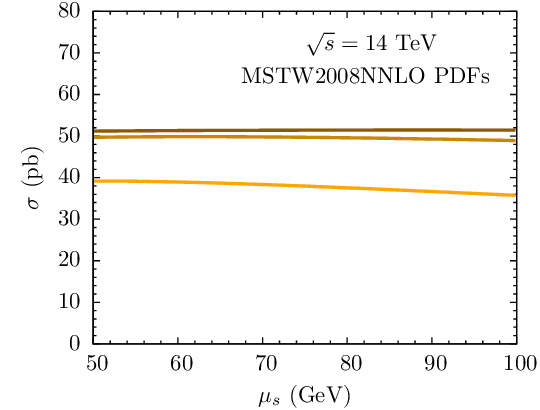}
\\[0.2cm]
\includegraphics[width=0.45\textwidth]{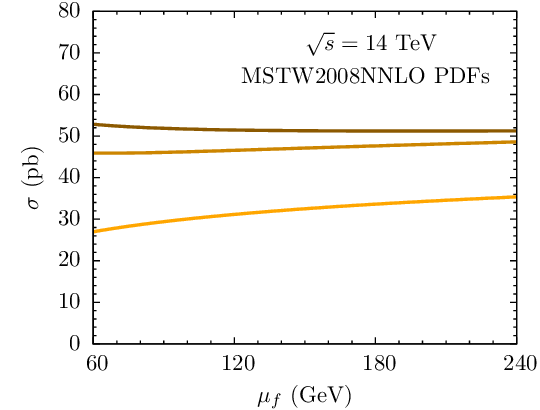}
\includegraphics[width=0.45\textwidth]{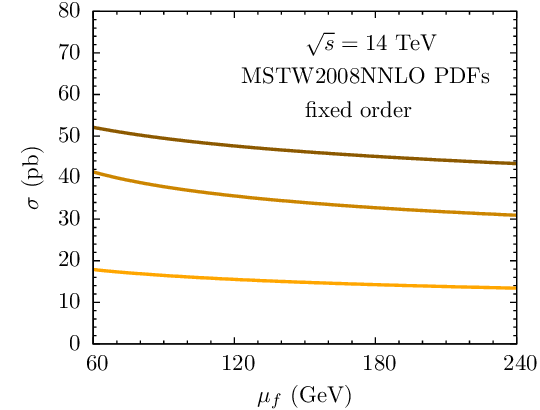}
\caption{\label{fig:scale1}
Dependence of the resummed cross section for Higgs-boson production at the LHC on the scales $\mu_h$ (upper left) and $\mu_s$ (upper right), and on the factorization scale $\mu_f$ (lower left), for $m_H=120$\,GeV. The darker curves correspond to higher orders in RG-improved perturbation theory. For comparison, we also show the dependence on $\mu_f$ in fixed-order perturbation theory (lower right). The corresponding curves for the Tevatron would look very similar except for the overall scale.}
\end{figure}

In our phenomenological analysis we vary the different scales independently about their default values and add up the corresponding variations of the cross section in quadrature. Specifically, we take $m_t/2<\mu_t<2m_t$, $(m_H/2)^2<-\mu_h^2<(2m_H)^2$, and $\mu_s^{\rm II}<\mu_s<\mu_s^{\rm I}$ for the matching scales, and $m_H/2<\mu_f<2m_H$ for the factorization scale. The effect of varying $\mu_t$ is so weak that we do not depict it in a plot. The $\mu_h$ and $\mu_s$ dependences of the cross sections obtained at different orders in RG-improved perturbation theory are shown in the upper panels of Figure~\ref{fig:scale1}. In both cases the scale dependence strongly decreases in higher orders and is essentially negligible already at NLO. In the lower panels of Figure~\ref{fig:scale1}, we compare the $\mu_f$ dependence of the combined cross section to that of the fixed-order cross section. Note that the $\mu_f$ dependence of the fixed-order cross section is not significantly improved when going from LO to NLO, and that a sizable $\mu_f$ dependence remains even at NNLO. On the other hand, the $\mu_f$ dependence after resummation is already small at LO, while the NLO and NNLO resummed cross sections only depend very weakly on the factorization scale.

\section{Predictions for the cross section} 
\label{sec:num}

\begin{figure}[t]
\centering
\includegraphics[width=0.49\textwidth]{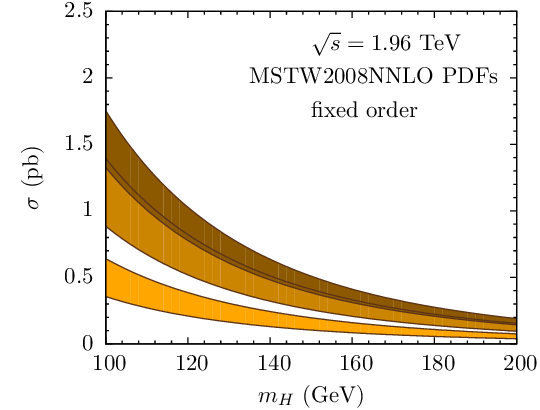}
\includegraphics[width=0.49\textwidth]{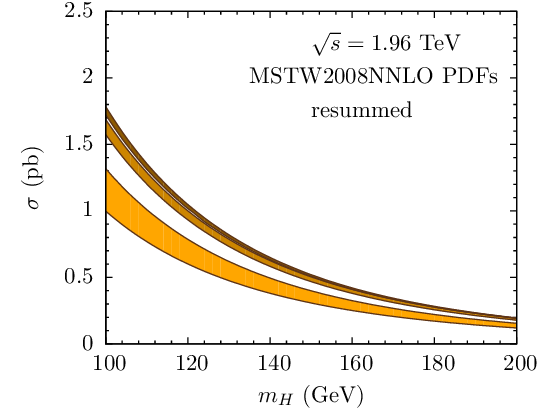}
\\
\includegraphics[width=0.48\textwidth]{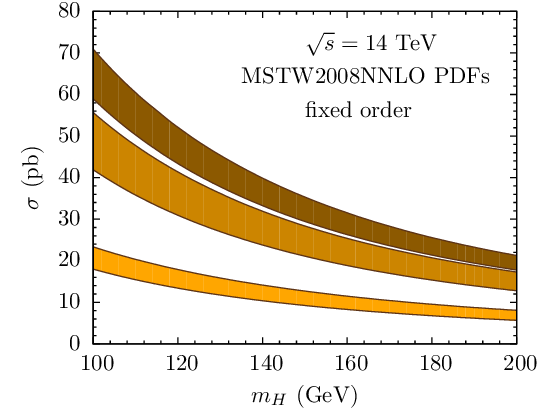}\hspace{2.5mm}
\includegraphics[width=0.48\textwidth]{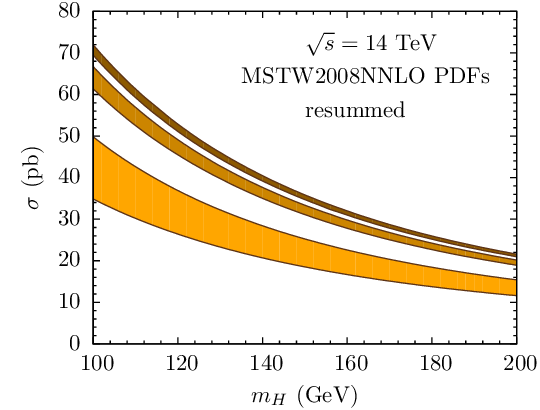}
\caption{\label{fig:cs}
The fixed-order (left) and RG-improved (right) cross-section predictions including perturbative uncertainty bands due to scale variations for the Tevatron (upper) and LHC (lower plots). Darker bands correspond to higher orders in perturbation theory.}
\end{figure}

\begin{figure}[t]
\centering
\includegraphics[width=0.49\textwidth]{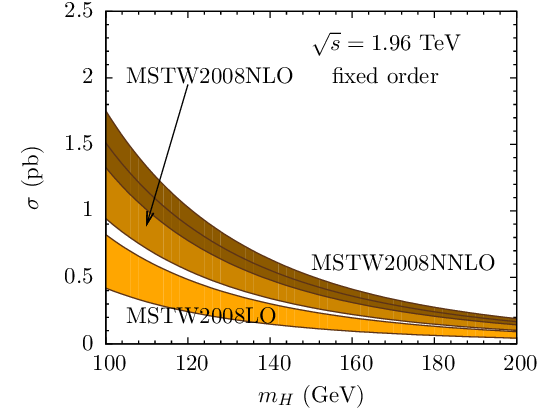}
\includegraphics[width=0.49\textwidth]{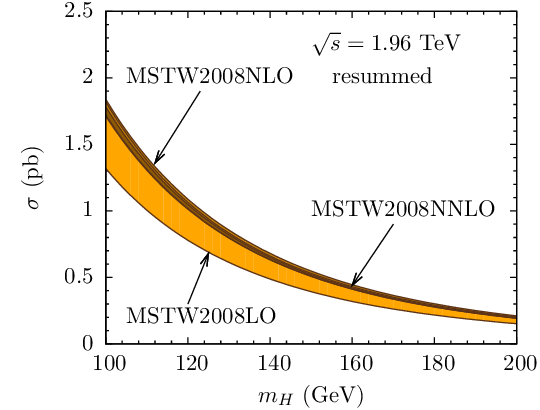}
\\
\includegraphics[width=0.48\textwidth]{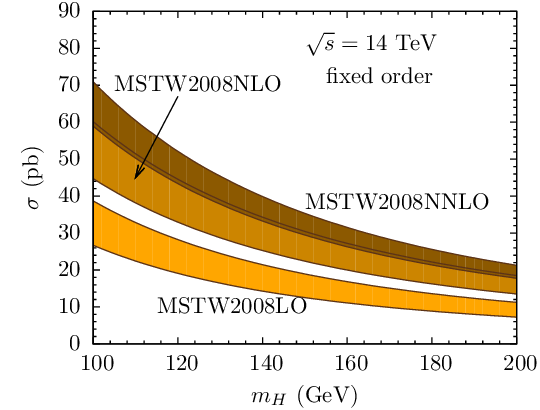}\hspace{2.5mm}
\includegraphics[width=0.48\textwidth]{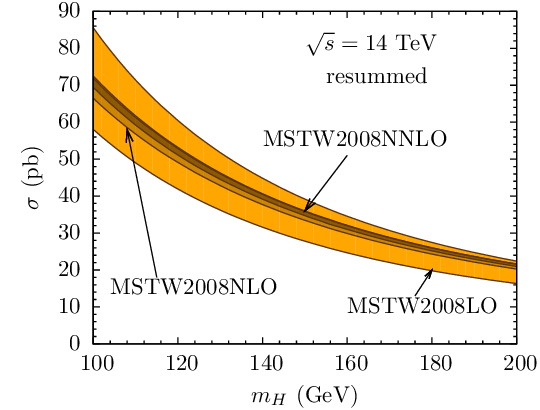}
\caption{\label{fig:csSwitch}
The fixed-order (left) and RG-improved (right) cross-section predictions including perturbative uncertainty bands due to scale variations for the Tevatron (upper) and LHC (lower plots). In contrast to Figure \ref{fig:cs}, different PDF sets are used according to the order of the calculation.
}
\end{figure}

We now present numerical results for the Higgs-boson production cross sections at the Tevatron and the LHC. To estimate the theoretical uncertainties we combine the various scale dependences as described in the previous section. The effect of the uncertainties in the PDFs is estimated by scanning over the 30 different sets provided by \cite{Martin:2009iq}. The uncertainty in the value of the running coupling $\alpha_s(m_Z^2)=0.1171\pm 0.0036$ introduces an additional error in the cross-section predictions of about $\pm 6\%$.
We compare our RG-improved results for the cross sections with those obtained in fixed-order perturbation theory. In the latter case we vary the factorization and renormalization scales together in the range $m_H/2<\mu_f<2m_H$. 

In Figure~\ref{fig:cs} we show the scale dependence of our predictions for the cross sections at different orders in perturbation theory. Note that we use the same PDFs (MSTW2008NNLO) in all cases, i.e., we do not switch to LO or NLO distribution functions in the lower-order results. This makes it easier to judge the actual size of the perturbative corrections to the hard-scattering kernels. The results obtained after RG improvement show significantly faster convergence and reduced scale dependence in higher orders. The NNLO resummed predictions have a perturbative uncertainty of less than 3\% for both the Tevatron and the LHC, while the scale dependence of the NNLO fixed-order results is approximately $\pm 15\%$ for the Tevatron and $\pm 10\%$ for the LHC. Numerical values for the cross section at NNLO are shown in Table~\ref{tab:results}. The first error accounts for scale variations, while the second one reflects the uncertainty in the PDFs. The additional uncertainty of $\pm 6\%$ due to the value of $\alpha_s(m_Z^2)$ is not shown explicitly. 
We emphasize that the effect of RG improvement is significant even at NNLO, where the resummed cross sections at the Tevatron and the LHC exceed the fixed-order predictions by about 13\% and 8\%, respectively (for $m_H=120$\,GeV). These differences are as important numerically as the differences between the NLO and NNLO resummed results. 

In Figure~\ref{fig:csSwitch}, we show for comparison the results obtained when the PDFs are switched according to the order of the calculation. When this is done, the higher-order bands obtained after RG improvement are fully contained in the lower-order ones and the $K$-factor is close to 1, in particular for the LHC.\footnote{For MRST2004 PDFs \cite{Martin:2006qz}, the $K$-factors after resummation are somewhat larger, $K\approx 1.3$ for the LHC, see \cite{Ahrens:2008qu}.} In fixed-order calculations it is customary to use PDFs extracted from a fit using predictions of the same order. Doing so absorbs universal higher-order corrections into the PDFs. Since resummed calculations contain contributions of arbitrarily high orders, the optimal PDF choice is less clear. If the same large higher-order corrections affect both the observable one tries to predict and the cross sections used to extract the PDFs, it would be quite problematic to perform a resummation in one case and not the other.  For our case, the relevant input quantity is the gluon PDF at low $x$, which is mostly determined by measurements of scaling violations in the DIS structure function, $\partial F_2(x,Q^2)/\partial Q^2$. The higher-order corrections associated with the analytic continuation of the time-like gluon form factor, which we resum, do not affect the DIS cross section, and so are not universal and cannot simply be absorbed into the PDFs. However, such corrections will be present for the Drell-Yan cross section, but smaller by a factor $C_F/C_A=4/9$. In Section \ref{sec:othercases}, we discuss their resummation for this process. Similar effects also appear in jet production processes, and it would be interesting to extend our method also to this case. Since jet production involves an interplay of time-like and space-like dynamics, the identification of the enhanced contributions will be more involved in this case.

While our numerical values include an uncertainty due to PDFs, it is important to note that  these uncertainties are estimated within the theoretical framework adopted when performing the PDF extraction. There are examples where newer PDF determinations have led to shifts which are larger than the quoted uncertainties. In particular, with the new MSTW2008NNLO PDFs the Higgs production cross section at the Tevatron is reduced by 10-15\% compared to MRST2006NNLO \cite{Martin:2007bv}, the previous-generation PDF set by the same collaboration.  Also, the MRST2006NNLO PDFs had lead to an increase of the production cross section at the LHC by 10\% compared to the result obtained using MRST2004NNLO \cite{Martin:2006qz}. For comparison, we provide in Table~\ref{tab:resultsCTEQ} results obtained using CTEQ6.6 PDFs \cite{Nadolsky:2008zw}. They are in good agreement with the results given in Table~\ref{tab:results}. Note, however, that the CTEQ6.6 PDFs are obtained from a fit to data using NLO cross sections only. Note also that the MSTW2008NNLO gluon PDF differs significantly from the one obtained by Alekhin et al.\ \cite{Alekhin:2005gq,Alekhin:2006zm}. 

\begin{table}
\caption{\label{tab:results}
Cross sections (in pb) for different Higgs masses at the Tevatron and the LHC, using MSTW2008NNLO PDFs. The first error accounts for scale variations, while the second one reflects the uncertainty in the PDFs.}
\vspace{0.2cm}
\centering
\begin{tabular}{ccccc}
\hline
\multirow{2}{*}{$m_H$ [GeV]} & \multicolumn{2}{c}{LHC}
 & \multicolumn{2}{c}{Tevatron} \\[-0.1cm]
 & fixed-order NNLO & resummed NNLO & fixed-order NNLO 
 & resummed NNLO \\ 
\hline \\[-0.4cm]
100 & $64.6^{+6.3+1.6}_{-5.7-2.1}$ & $69.9^{+2.1+1.8}_{-0.4-2.3}$ & $1.536^{+0.216+0.082}_{-0.210-0.089}$ & $1.735^{+0.049+0.089}_{-0.013-0.096}$
\\
120 & $47.6^{+4.5+1.1}_{-4.2-1.5}$ & $51.4^{+1.4+1.2}_{-0.3-1.6}$ & $0.901^{+0.126+0.056}_{-0.124-0.060}$ & $1.022^{+0.025+0.061}_{-0.005-0.065}$
\\
140 & $36.5^{+3.4+0.9}_{-3.2-1.1}$ & $39.4^{+1.1+0.9}_{-0.2-1.2}$ & $0.559^{+0.078+0.040}_{-0.077-0.043}$ & $0.636^{+0.013+0.055}_{-0.004-0.046}$
\\
160 & $28.8^{+2.7+0.7}_{-2.5-0.8}$ & $31.2^{+0.8+0.7}_{-0.2-0.9}$ & $0.361^{+0.050+0.029}_{-0.050-0.031}$ & $0.413^{+0.007+0.032}_{-0.002-0.034}$
\\
180 & $23.4^{+2.1+0.5}_{-2.1-0.7}$ & $25.3^{+0.6+0.6}_{-0.2-0.7}$ & $0.242^{+0.034+0.022}_{-0.034-0.023}$ & $0.278^{+0.004+0.024}_{-0.001-0.025}$
\\
200 & $19.5^{+1.8+0.5}_{-1.7-0.6}$ & $21.1^{+0.5+0.5}_{-0.1-0.6}$ & $0.167^{+0.023+0.016}_{-0.024-0.017}$ & $0.193^{+0.002+0.018}_{-0.001-0.019}$
\\
220 & $16.6^{+1.5+0.4}_{-1.5-0.5}$ & $17.9^{+0.4+0.4}_{-0.1-0.5}$ & $0.118^{+0.016+0.013}_{-0.017-0.013}$ & $0.138^{+0.002+0.014}_{-0.001-0.015}$
\\
240 & $14.4^{+1.3+0.4}_{-1.3-0.4}$ & $15.5^{+0.3+0.4}_{-0.1-0.5}$ & $0.086^{+0.012+0.010}_{-0.012-0.010}$ & $0.100^{+0.002+0.011}_{-0.000-0.011}$
\\
260 & $12.7^{+1.1+0.3}_{-1.1-0.4}$ & $13.8^{+0.3+0.4}_{-0.1-0.4}$ & $0.064^{+0.009+0.008}_{-0.009-0.008}$ & $0.075^{+0.002+0.009}_{-0.000-0.009}$
\\
280 & $11.5^{+1.0+0.3}_{-1.0-0.4}$ & $12.4^{+0.3+0.3}_{-0.1-0.4}$ & $0.048^{+0.007+0.006}_{-0.007-0.007}$ & $0.057^{+0.001+0.007}_{-0.000-0.007}$
\\
300 & $10.6^{+0.9+0.3}_{-0.9-0.3}$ & $11.5^{+0.2+0.3}_{-0.1-0.4}$ & $0.038^{+0.005+0.005}_{-0.006-0.005}$ & $0.045^{+0.001+0.006}_{-0.000-0.006}$
\\[0.2cm]
\hline
\end{tabular}
\end{table}

\begin{table}
\caption{\label{tab:resultsCTEQ}
Cross sections (in pb) for different Higgs masses at the Tevatron and the LHC, using CTEQ6.6 PDFs. The first error accounts for scale variations, while the second one reflects the uncertainty in the PDFs.}
\vspace{0.2cm}
\centering
\begin{tabular}{ccccc}
\hline
\multirow{2}{*}{$m_H$ [GeV]} & \multicolumn{2}{c}{LHC}
 & \multicolumn{2}{c}{Tevatron} \\[-0.1cm]
 & fixed-order NNLO & resummed NNLO & fixed-order NNLO 
 & resummed NNLO \\ 
\hline \\[-0.4cm]
100 & $65.9^{+6.6+2.5}_{-6.0-3.0}$ & $71.4^{+2.2+2.8}_{-0.6-3.3}$ & $1.516^{+0.217+0.074}_{-0.209-0.076}$ & $1.716^{+0.048+0.078}_{-0.009-0.081}$
\\
120 & $48.2^{+4.7+1.7}_{-4.3-2.0}$ & $52.3^{+1.5+1.8}_{-0.3-2.2}$ & $0.890^{+0.126+0.058}_{-0.123-0.056}$ & $1.011^{+0.024+0.061}_{-0.005-0.060}$
\\
140 & $36.8^{+3.5+1.2}_{-3.3-1.5}$ & $39.9^{+1.1+1.3}_{-0.2-1.6}$ & $0.553^{+0.077+0.046}_{-0.077-0.043}$ & $0.631^{+0.013+0.049}_{-0.002-0.046}$
\\
160 & $29.0^{+2.8+0.8}_{-2.6-1.0}$ & $31.4^{+0.8+0.9}_{-0.2-1.1}$ & $0.359^{+0.050+0.037}_{-0.050-0.033}$ & $0.411^{+0.007+0.039}_{-0.002-0.036}$
\\
180 & $23.5^{+2.2+0.6}_{-2.1-0.8}$ & $25.5^{+0.6+0.7}_{-0.1-0.9}$ & $0.242^{+0.034+0.030}_{-0.034-0.026}$ & $0.278^{+0.004+0.032}_{-0.001-0.028}$
\\
200 & $19.5^{+1.8+0.5}_{-1.8-0.6}$ & $21.1^{+0.5+0.6}_{-0.1-0.7}$ & $0.168^{+0.023+0.024}_{-0.024-0.021}$ & $0.194^{+0.002+0.026}_{-0.001-0.023}$
\\
220 & $16.5^{+1.5+0.4}_{-1.5-0.5}$ & $17.9^{+0.4+0.5}_{-0.1-0.5}$ & $0.120^{+0.017+0.020}_{-0.017-0.017}$ & $0.140^{+0.002+0.021}_{-0.001-0.018}$
\\
240 & $14.3^{+1.3+0.4}_{-1.3-0.4}$ & $15.5^{+0.4+0.4}_{-0.1-0.4}$ & $0.088^{+0.012+0.016}_{-0.012-0.014}$ & $0.103^{+0.002+0.018}_{-0.000-0.015}$
\\
260 & $12.6^{+1.1+0.3}_{-1.1-0.3}$ & $13.7^{+0.3+0.3}_{-0.1-0.4}$ & $0.066^{+0.009+0.014}_{-0.009-0.011}$ & $0.078^{+0.002+0.015}_{-0.000-0.012}$
\\
280 & $11.4^{+1.0+0.3}_{-1.0-0.3}$ & $12.4^{+0.3+0.3}_{-0.1-0.4}$ & $0.051^{+0.007+0.012}_{-0.007-0.010}$ & $0.060^{+0.001+0.013}_{-0.000-0.011}$
\\
300 & $10.5^{+0.9+0.3}_{-1.0-0.3}$ & $11.4^{+0.2+0.3}_{-0.1-0.3}$ & $0.040^{+0.006+0.010}_{-0.006-0.008}$ & $0.048^{+0.001+0.011}_{-0.000-0.009}$
\\[0.2cm]
\hline
\end{tabular}
\end{table}

In our predictions we resum logarithmic terms near the partonic threshold as well as the $\pi^2$-enhanced terms contained in the hard matching coefficient $H$ in (\ref{Cfact}). It is simple to disentangle the two effects: choosing $\mu_h^2=m_H^2$ instead of $\mu_h^2=-m_H^2$ switches off the resummation of the $\pi^2$ terms. With this choice our results are equivalent to what is obtained in standard soft-gluon resummation, albeit performed in momentum space instead of Mellin moment space.  As seen in Table \ref{tab:ipi}, the main effect of RG improvement is the resummation of the $\pi^2$-enhanced terms contained in the hard matching coefficient $H$ in (\ref{Cfact}). The predictions for the resummed cross section obtained without resummation of the $\pi^2$-enhanced terms are quite close to the fixed-order results. This shows once again that soft-gluon resummation is {\em not\/} an important effect in the case of Higgs-boson production at Tevatron or LHC energies. It confirms our theoretical argument given in conjunction with relation (\ref{Born}) and is also in line with our finding that the optimal value of the soft scale $\mu_s$ is not much lower than hard scale set by the Higgs mass. 

Compared to the numerical results obtained in \cite{Catani:2003zt} using the traditional moment-space formalism, we find smaller threshold resummation effects. Part of the difference is simply due to the fact that we use the values obtained with our default scale choices as our default values. As a consequence our uncertainty bands are quite asymmetric. In contrast, \cite{Catani:2003zt} uses the central values of the bands as the default values. However, even after taking this trivial difference into account, we observe that the resummation effects found in \cite{Catani:2003zt} are larger than those we find. While compatible within the assigned uncertainties, we find that the resummation effects obtained using the moment formalism of \cite{Catani:2003zt} are about twice as large as those we find in momentum space. Specifically, evaluating the NNLL moment-space result of \cite{Catani:2003zt} with MSTW2008NNLO PDFs, and matching to the NNLO fixed-order result, we obtain $\sigma=50.4\,{\rm pb}$ at the LHC for $m_H=120\,{\rm GeV}$, compared to our result $\sigma=48.5\,{\rm pb}$ and the fixed-order value of $47.6\,{\rm pb}$. 

There are three differences between our calculation and what was done in \cite{Catani:2003zt}: (i) we go one order higher in logarithmic accuracy, (ii) instead of the scale choice $\mu_s\sim m_H/N$ inherent in the moment-space formalism we set the scale as discussed in Section \ref{sec:scales}, and (iii) while the two formalisms are equivalent in the threshold region, the power-suppressed terms differ between the two formulations. Neither the additional higher-log contributions nor the scale-setting prescription can account for the difference. To compare the two scale-setting prescriptions, we have evaluated the effective-theory moment-space result (\ref{CN}) both with $\mu_s\sim m_H/N$ and with our choice of the soft scale and find that the difference is small. Using our default choice for the soft scale, the effective-theory moment-space result is $50.1\,{\rm pb}$, very close to what is obtained in the traditional framework. The difference thus arises from power corrections suppressed by $(1-z)$ or $1/N$, respectively. If we use the same scale setting in the moment-space formula (\ref{CN}) and momentum-space expression (\ref{eq:cc}), then the difference between the two formulations amounts to an overall factor of $\sqrt{z}$, see (\ref{Mellininv}). To check that this factor indeed accounts for the difference, we have multiplied our momentum space formula (\ref{eq:cc}) by $\sqrt{z}$. After adjusting the matching corrections, we find $\sigma=49.9{\rm pb}$ instead of $48.5{\rm pb}$.\footnote{Note that the enhancement is very counter-intuitive: the procedure amounts to multiplying the parton luminosity $f\hspace{-0.13cm}f_{gg}(\tau/z)$ by $\sqrt{z}$ which is smaller than 1 over the entire integration range and nevertheless leads a larger result. The enhancement arises because the kernel is a distribution and its plus-distribution part is sensitive to the derivative of $\sqrt{z} f\hspace{-0.13cm}f_{gg}(\tau/z)$.} The factor $\sqrt{z}$  appears artificial, since it does not occur in the fixed-order expressions. On the other hand, with this factor included, the singular terms are larger and amount numerically to 96\% of the full NNLO result (without this factor, they amount to 86\%). As stressed above, the threshold dominance is observed numerically but not enforced parametrically. For this reason, equivalent definitions of the leading contribution can lead to somewhat different results. 

\begin{table}
\caption{\label{tab:ipi}
Cross sections (in pb) for $m_H=120$\,GeV. We compare fixed-order results (first column) with RG-improved results (remaining three columns) corresponding to standard soft-gluon resummation with $\mu_h^2=+m_H^2$, resummation of $\pi^2$-enhanced terms ($\mu_h^2=-m_H^2$) only, and the combination of both. The uncertainties are due to scale and PDF variation.}
\vspace{0.2cm}
\centering
\begin{tabular}{crcccc}
\hline
 & & fixed order & threshold & $\pi^2$-enhanced 
 & $\mbox{threshold} + \pi^2$ \\ 
\hline \\[-0.4cm]
      & LO & $15.5^{+2.4+0.4}_{-2.1-0.5}$ & $17.8^{+3.3+0.4}_{-2.7-0.6}$ &
      $27.1^{+4.0+0.6}_{-3.8-0.8}$ & $31.2^{+5.7+0.8}_{-4.8-1.0}$
      \\
      LHC & NLO & $35.5^{+5.9+0.8}_{-4.6-1.1}$ & $37.7^{+3.6+0.9}_{-1.2-1.2}$ &
      $45.0^{+3.0+1.1}_{-3.3-1.4}$ & $46.6^{+2.5+1.1}_{-1.1-1.5}$
      \\
      & NNLO & $47.6^{+4.5+1.1}_{-4.2-1.5}$ & $48.5^{+2.5+1.2}_{-0.5-1.5}$ &
      $51.4^{+1.7+1.2}_{-1.6-1.6}$ & $51.4^{+1.4+1.2}_{-0.3-1.6}$
\\[0.2cm] 
\hline \\[-0.4cm]
     & LO & $0.281^{+0.105+0.018}_{-0.071-0.019}$ &
      $0.389^{+0.062+0.023}_{-0.046-0.024}$ &
      $0.491^{+0.180+0.031}_{-0.127-0.033}$ & $0.681^{+0.105+0.040}_{-0.080-0.042}$
      \\
      Tevatron & NLO & $0.650^{+0.172+0.041}_{-0.131-0.044}$ &
      $0.764^{+0.077+0.045}_{-0.026-0.048}$ &
      $0.855^{+0.125+0.053}_{-0.130-0.056}$ & $0.954^{+0.046+0.055}_{-0.022-0.059}$
      \\
      & NNLO & $0.901^{+0.126+0.056}_{-0.124-0.060}$ &
      $0.961^{+0.048+0.058}_{-0.012-0.062}$ &
      $1.003^{+0.051+0.061}_{-0.074-0.065}$ & $1.022^{+0.025+0.061}_{-0.005-0.065}$
\\[0.2cm] \hline
\end{tabular}
\end{table}

To conclude our discussion, let us briefly discuss the case of Higgs production with a jet veto, i.e. the cross section for the production of the Higgs boson and QCD radiation with $p_T \equiv |\vec{p}_T|< p_T^{\rm veto}$. Such a veto reduces the background to $H\to W^+W^- \to l^+l^-\nu\bar\nu$ from $t\bar{t}$-production with subsequent $t\to b\ell^+\bar\nu$ decay. It was observed that the $K$-factor for Higgs production gets reduced when such a cut is imposed \cite{Catani:2001cr,Anastasiou:2004xq,Anastasiou:2007mz}. For example, at the LHC with $p_T^{\rm veto}=30\,{\rm GeV}$ and $m_H=165\,{\rm GeV}$, reference \cite{Anastasiou:2007mz} finds $K_{\rm NLO}\approx K_{\rm NNLO}\approx1.4$, while the inclusive $K$-factors are $K_{\rm NLO}\approx 1.8 $ and $K_{\rm NNLO}\approx 2.2$ (using MRST2004 PDFs). 
This implies that a significant portion of the perturbative corrections comes from the region $p_T> 30\,{\rm GeV}$. Since the cut only excludes hard radiation, the threshold region should not be affected by it. At first sight, these large corrections from hard radiation seem difficult to reconcile with our finding that roughly 90\% of the NLO and NNLO total cross sections arise from the partonic threshold terms. 

More formally, for a given value of $z$ we have $p_T< \frac{M_H (1-z)}{2\sqrt{z}}$, which shows that the radiation from the threshold region $z\to 1$ has small $p_T$. The veto $p_T< p_T^{\rm veto}$ therefore does not constrain the radiation above a certain value of $z>z_0$. On the other hand, the converse is not true: small $p_T$ does not imply that the radiation is soft. To isolate the amount of soft radiation present for a given $p_T$ cut, we have imposed a cut $z>z_0$ in the integration over the leading singular terms. For $p_T= 30\, {\rm GeV}$ and $M_H=165\, {\rm GeV}$, we have $z_0=0.7$. We find that evaluating the threshold terms (\ref{fixedorder}) with such a cut leaves them essentially unchanged. 
So we are forced to conclude that the cross section for $p_T< p_T^{\rm veto}$ is smaller than the contribution from the threshold terms alone, which implies that hard radiation with small $p_T$, which comprises radiation with a small angle with respect to the beam, gives a sizable negative correction to the cross section. 

It is difficult to draw any conclusions from the above discussion, other than the trivial statement that also processes involving hard radiation can receive large higher-order corrections. To understand whether there is a physics reason for the observed reduction of the $K$-factors in the presence of a jet veto, it would be interesting to analyze Higgs production in association with a high-$p_T$ jet in the effective theory. This process is mediated by operators involving additional collinear fields for the partons inside the jet. The corresponding hard function will contain both space-like and time-like Sudakov double logarithms, which will need to be disentangled, before the enhanced contributions can be resummed and compared to the enhanced contributions affecting the total rate.

\section{\boldmath RG-improvement for other time-like processes}

Having discussed Higgs-boson production in detail, we now briefly explore the effect of choosing a time-like renormalization point $\mu^2<0$ for other processes. Our treatment applies immediately to Drell-Yan production, but the numerical effects are less dramatic than for Higgs-boson production, as we pointed out in \cite{Ahrens:2008qu}. In addition, it is interesting to compare these production processes to inclusive decays such as $e^+ e^-\to\mbox{hadrons}$, $\tau\to\nu_\tau+\mbox{hadrons}$, or hadronic Higgs-boson decay. For inclusive decay rates Sudakov double logarithms and the associated $\pi^2$-enhanced terms are absent, since they cancel between real and virtual corrections by virtue of the KLN theorem \cite{Kinoshita:1962ur,Lee:1964is}. As a consequence, the effects of choosing $\mu^2<0$ are small unless the characteristic momentum scale is quite low. 

\subsection{Drell-Yan process}
\label{sec:othercases}

Near the partonic threshold, the Drell-Yan cross section factors into a hard and a soft function, and threshold resummation proceeds in complete analogy to the Higgs case \cite{Becher:2007ty}. Instead of the scalar two-gluon operator (\ref{Heff}), Drell-Yan production is mediated by the electromagnetic current $\bar q\gamma_\mu q$. The hard function is given by the renormalized on-shell vector form factor $C_V(-q^2,\mu_h^2)$. The same hard function also appears in deep-inelastic scattering, but evaluated at space-like momentum transfer. The analytic continuation of the form factor to the time-like region produces $\pi^2$ terms, which were resummed in \cite{Parisi:1979xd,Magnea:1990zb,Bakulev:2000uh,Eynck:2003fn}. The Drell-Yan case can be treated in exactly the same way as Higgs-boson production. The quantity $C_V$ fulfills a RG equation of the same form as (\ref{CSevol}) for $C_S$, however the relevant cusp anomalous dimension $\Gamma^F_{\rm cusp}$ is smaller by a factor $C_F/C_A=4/9$. The resummation effects are thus smaller than in the case of Higgs-boson production and have the form $\exp(C_F\alpha_s\pi/2)$ at leading order, see (\ref{Ureexpanded}). Comparing the expansion of the hard function at time-like and space-like renormalization points, we find
\begin{equation}
\begin{aligned}
   |C_V(-q^2,q^2)|^2
   &= 1 + 0.0845 + 0.0292 + \dots \,, \\
   |C_V(-q^2,-q^2)|^2
   &= 1 - 0.1451 - 0.0012 + \dots \,.
\end{aligned}
\end{equation}
The two-loop correction is reduced for $\mu^2=-q^2$, however, at one-loop order the correction increases since there is a partial cancellation between the $\pi^2$-enhanced terms and the constant piece for $\mu_h^2>0$. Numerical results for the resummed Drell-Yan cross section are given in Table~\ref{tab:drellyanNum} for the case of E866/NuSea \cite{Webb:2003ps}, i.e., proton-proton collisions at $\sqrt{s}=38.76$\,GeV and $q^2=(8\,{\rm GeV})^2$. The scales $|\mu_h|$, $\mu_s$, and $\mu_f$ in the resummation formula for Drell-Yan production have been chosen as in \cite{Becher:2007ty}, and MRST2004 PDFs were used \cite{Martin:2006qz}. The numbers in the table include the matching to fixed-order perturbation theory at the corresponding order. At NNLO, the difference between ordinary threshold resummation with $\mu_h^2>0$ and the combined resummation with $\mu_h^2<0$ is about 4\%, significant because of the large $\alpha_s$ value at such low energies. Convergence is similar in both cases, with negative instead of positive corrections for $\mu_h^2<0$. The numbers obtained in the two cases do not agree within their respective scale uncertainties. They would be compatible if the hard scale would be varied up and down by a factor 2 as we do in the present paper. However, a smaller variation $\sqrt{q^2}<\mu_h<2\sqrt{q^2}$ was used in \cite{Becher:2007ty}. In view of the disagreement, a variation by the conventional factor of 2 seems more appropriate.

\begin{table}
\caption{\label{tab:drellyanNum}
Predictions for the Drell-Yan cross section $d\sigma/dq^2$ at $\sqrt{s}=38.76$\,GeV for an invariant mass of $\sqrt{q^2}=8$\,GeV of the lepton pair. Units are pb/GeV$^2$.}
\vspace{0.2cm}
\centering
\begin{tabular}{rccc}
\hline
 & fixed order & threshold & $\mbox{threshold} + \pi^2$ \\ 
\hline \\[-0.4cm]
LO & $0.299_{\,-0.040}^{\,+0.051}$ & $0.436_{\,-0.071}^{\,+0.062}$
 & $0.700_{\,-0.106}^{\,+0.091}$ \\
NLO & $0.449_{\,-0.041}^{\,+0.051}$ & $0.493_{\,-0.014}^{\,+0.011}$ 
 & $0.559_{\,-0.035}^{\,+0.014}$ \\
NNLO & $0.505_{\,-0.025}^{\,+0.021}$ & $0.512_{\,-0.004}^{\,+0.002}$ 
 & $0.534_{\,-0.006}^{\,+0.009}$
\\[0.2cm] \hline
\end{tabular}
\end{table}

The main goal of the E866/NuSea experiment was to provide a determination of the sea-quark PDFs of the proton. The resummation of $\pi^2$ terms would affect this determination in the form of an overall normalization factor. The absolute normalization of the cross section is also interesting at higher energies, e.g.\ for using the Drell-Yan process to monitor the luminosity at the LHC. However, at NNLO the difference between the two scale-setting prescriptions scales as $\alpha_s^3$ and would thus be four times smaller at $\sqrt{q^2}=m_Z$ than at $\sqrt{q^2}=8$\,GeV, and completely negligible for higher-mass Drell-Yan pairs.

\subsection{\boldmath $e^+ e^-\to\mbox{hadrons}$ and hadronic $\tau$ decays\unboldmath}

The total $e^+ e^-\to\mbox{hadrons}$ cross section satisfies the relation
\begin{equation}\label{Ree}
   R(s) = \frac{\sigma(e^+ e^-\to\mbox{hadrons})}%
               {\sigma(e^+ e^-\to\mu^+\mu^-)}
   = N_c\,\Big( \sum_q\,e_q^2 \Big)\,4\pi\,
   \mbox{Im}\Pi_{q\bar q}(-s+i\epsilon) \,,
\end{equation}
where the sum extends over all quark flavors with $2m_q<\sqrt{s}$, and the current-current vacuum correlator $\Pi_{q\bar q}(Q^2)$ is related to the Adler $D$-function as
\begin{equation}\label{Ddef}
   D(Q^2) = 4\pi^2\,\frac{d\Pi_{q\bar q}(Q^2)}{d\ln Q^2} \,.
\end{equation}
The quantity $\Pi_{q\bar q}(-q^2+i\epsilon)$ denotes its analytic continuation to the region of time-like momentum transfer. For simplicity, we neglect the masses of the light quarks and assume that $s$ is far away from quark thresholds.

The Adler function in massless QCD is RG invariant, implying that its evolution equation $dD(Q^2)/d\ln\mu=0$ is trivially free of the cusp contributions associated with Sudakov logarithms and the large $\pi^2$ terms encountered in the case of Higgs-boson and Drell-Yan production. It follows that the perturbative expansion of the Adler function can be written as
\begin{equation}\label{Dexp}
   D(Q^2) = 1 + \sum_{n=1}^\infty d_n 
   \left( \frac{\alpha_s(Q^2)}{4\pi} \right)^n ,
\end{equation}
where the expansion coefficients $d_n$ are pure numbers, independent of the renormalization scale. Explicitly we have (setting $N_c=3$) \cite{Chetyrkin:1979bj,Dine:1979qh,Celmaster:1980ji}
\begin{equation}
   d_1 = 4 \,, \qquad
   d_2 = \left( 22 - 16\zeta_3 \right) \beta_0 + \frac43
    \approx 2.77\beta_0 + 1.33 \,.
\end{equation}
In contrast to the case of Higgs-boson production, $\pi^2$-enhanced perturbative corrections enter the Adler function and the $e^+ e^-\to\mbox{hadrons}$ cross section only at ${\cal O}(\alpha_s^3)$ and higher. Reexpressing the QCD coupling in terms of $\alpha_s(\mu^2)$, integrating relation (\ref{Ddef}), and analytically continuing to the time-like region, $Q^2\to-s+i\epsilon$, we obtain
\begin{equation}\label{Rexp}
   4\pi^2\,\mbox{Im}\Pi_{q\bar q}(-s+i\epsilon)
   = 1 + d_1\,\frac{\alpha_s(s)}{4\pi}
    + d_2 \left( \frac{\alpha_s(s)}{4\pi} \right)^2 + \dots \,.
\end{equation}
This formula is routinely used in the calculation of the total cross section. Differences between the perturbative series in (\ref{Dexp}) and (\ref{Rexp}) arise starting at ${\cal O}(\alpha_s^3)$.

The RG-improved expression for the imaginary part of the correlator is obtained by expanding the result for the current-current correlator $\Pi_{q\bar q}(-s+i\epsilon)$ in powers of $\alpha_s(-s+i\epsilon)$, i.e., using perturbation theory at time-like momentum transfer \cite{Radyushkin:1982kg,Pennington:1981cw}. Integrating relation (\ref{Ddef}), we have
\begin{equation}
\begin{split}
   &4\pi\,\Pi_{q\bar q}(-s+i\epsilon)
    = \frac{2}{\pi} \int\limits^{\alpha_s(-s+i\epsilon)}\!\!d\alpha\,
    \frac{D(\alpha)}{\beta(\alpha)} + \mbox{const.} \\
   &= \frac{1}{\pi} \ln(-s+i\epsilon)
    - \frac{1}{\pi\beta_0} \left[ d_1 \ln\alpha_s(-s+i\epsilon)
    + \left( d_2 - \frac{d_1\beta_1}{\beta_0} \right)\,
    \frac{\alpha_s(-s+i\epsilon)}{4\pi} + \dots \right] 
    + \mbox{const.,}
\end{split}
\end{equation}
where $\beta_1/\beta_0=(102-38n_f/3)/(11-2n_f/3)$. Eliminating the time-like coupling in favor of the space-like one using (\ref{asNLO}), we obtain
\begin{equation}\label{eq40}
\begin{split}
   4\pi\,\mbox{Im}\Pi_{q\bar q}(-s+i\epsilon)
   &= 1 + d_1\,\frac{\alpha_s(s)}{4\pi}\,\frac{\arctan(a_s)}{a_s} \\
   &\quad\mbox{}+ \frac{d_2}{1+a_s^2}
    \left( \frac{\alpha_s(s)}{4\pi} \right)^2
    \left[ 1 + \frac{d_1\beta_1}{d_2\beta_0}
    \left( \frac{\arctan(a_s)}{a_s} - \frac{\ln(1+a_s^2)}{2} - 1 
    \right) \right] + \dots \,,
\end{split}
\end{equation}
where $a_s\equiv\beta_0\alpha_s(s)/4$. This formula was first derived in \cite{Radyushkin:1982kg}. It is the RG-improved version of (\ref{Rexp}), which should be used whenever $a_s$ is an ${\cal O}(1)$ parameter. For example, with $a_s=0.7$ as appropriate for $s=m_\tau^2$ we find
\begin{equation}
   4\pi\,\mbox{Im}\Pi_{q\bar q}(-s+i\epsilon)
   = 1 + 0.872\,d_1\,\frac{\alpha_s(s)}{4\pi}
   + d_2 \left( \frac{\alpha_s(s)}{4\pi} \right)^2
   \left( 0.671 - 0.219\,\frac{d_1\beta_1}{d_2\beta_0} \right)
   + \dots \,,
\end{equation}
which for $n_f=4$ yields a reduction of the two-loop coefficient by a factor of 0.45. For higher values of $s$ corresponding to weak-scale processes, on the other hand, the modifications with respect to (\ref{Rexp}) are very small. 

Relation (\ref{eq40}) shows that $\pi^2$-enhanced corrections to the $e^+ e^-\to\mbox{hadrons}$ cross sections arise first at ${\cal O}(\alpha_s^3)$ in fixed-order perturbation theory. The leading term results from the expansion of the $\arctan(a_s)/a_s$ factor in the first line and yields $(-\pi^2\beta_0^2/48)(\alpha_s/\pi)^3\approx -14.3\,(\alpha_s/\pi)^3$ for $n_f=4$, which is a rather large correction. For this reason, it was argued in \cite{Radyushkin:1982kg} that for time-like processes $\bar\alpha_s(s)\equiv(4/\beta_0)\arctan(a_s)$ is a better expansion parameter than $\alpha_s(s)$. As an alternative choice, the authors of \cite{Pennington:1981cw} suggested to use $|\alpha_s(-s)|\approx\alpha_s(s)/\sqrt{1+a_s^2}$ for the expansion parameter. Note that both quantities have the property that they ``freeze'' in the infrared.

Our RG-improved result (\ref{eq40}) coincides with the expression for the cross section obtained using contour-improved perturbation theory \cite{Pivovarov:1991rh,Le Diberder:1992te}. The analytic properties of the Adler function imply the relation
\begin{equation}\label{Dcircle}
   4\pi\,\mbox{Im}\Pi_{q\bar q}(-s+i\epsilon)
   = \frac{1}{2\pi i} \oint\limits_{|s'|=s}\!\frac{ds'}{s'}\,D(-s') 
   = \frac{1}{2\pi} \int_{-\pi}^\pi\!d\varphi\,D(e^{i\varphi} s) \,.
\end{equation}
Inserting here the expansion (\ref{Dexp}) and using the generalization
\begin{equation}\label{ascomplex}
   \frac{\alpha_s(\mu^2)}{\alpha_s(e^{i\varphi}\mu^2)}
   = 1 + ia_\varphi 
   + \frac{\beta_1}{\beta_0}\,\frac{\alpha_s(\mu^2)}{4\pi}\, 
   \ln(1+ia_\varphi) + {\cal O}(\alpha_s^2) \,,
\end{equation}
of relation (\ref{asNLO}), where $a_\varphi=(\varphi/\pi)\,\beta_0\alpha_s(\mu^2)/4$, we readily recover (\ref{eq40}). The relevant contour integrals are evaluated in Appendix~\ref{app2}. Weighted contour integrals over $D$ functions can also be used to calculate the total hadronic decay rate of the $\tau$ lepton and, more generally, moments of the corresponding spectral functions accessible in $\tau$ decay \cite{Le Diberder:1992fr,Neubert:1995gd}. The equivalence of our approach with contour-improved perturbation theory extends to these cases as well.

\subsection{Hadronic Higgs-boson decay}

The total $H\to gg$ decay rate may be written as
\begin{equation}
\begin{split}
   \Gamma(H\to gg)
   &= \frac{G_F m_H^3}{36\pi^3\sqrt2}\,K \,, \\
   K &= \alpha_s^2(\mu^2) \left[ C_t(m_t^2,\mu^2) \right]^2\,
    \frac{\pi}{2m_H^4}\,\mbox{Im}\Pi_{gg}(-m_H^2+i\epsilon,\mu^2) \,,
\end{split}
\end{equation}
where the correlator $\Pi_{gg}$ is defined as in \cite{Baikov:2006ch}, and the leading-order $K$-factor is $K_{\rm LO}=\alpha_s^2(\mu^2)$. We can use the RG-improved expression (\ref{Ctsol}) for the matching coefficient $C_t(m_t^2,\mu^2)$ to rewrite the $K$-factor in the form
\begin{equation}
   K = \left[ \frac{C_t^2(m_t^2,\mu_t^2)}{b^2(\mu_t^2)} \right]
   \left[ \alpha_s^2(\mu^2)\,b^2(\mu^2)\,
    \frac{\pi}{2m_H^4}\,\mbox{Im}\Pi_{gg}(-m_H^2+i\epsilon,\mu^2)
    \right]
   \equiv K_t(m_t^2)\,K_H(m_H^2) \,,
\end{equation}
where we have introduced the function
\begin{equation}
   b(\mu^2) 
   = \frac{\beta(\alpha_s(\mu^2))/\alpha_s^2(\mu^2)}{-\beta_0/2\pi}
   = 1 + \frac{\beta_1}{\beta_0}\,\frac{\alpha_s(\mu^2)}{4\pi}
    + \frac{\beta_2}{\beta_0}
    \left( \frac{\alpha_s(\mu^2)}{4\pi} \right)^2 + \dots \,.
\end{equation}

The $K$-factor is independent of the scales $\mu_t$ and $\mu$ and factorizes into a product of two RG-invariant quantities depending on the scales $m_t$ and $m_H$. Evaluating the first factor with $\mu_t=m_t$, we obtain (setting $N_c=3$)
\begin{equation}
   K_t(m_t^2) = \frac{C_t^2(m_t^2,\mu_t^2)}{b^2(\mu_t^2)}
   = 1 + \left( 22 - \frac{2\beta_1}{\beta_0} \right) 
   \frac{\alpha_s(m_t^2)}{4\pi} + {\cal O}(\alpha_s^2) \,.
\end{equation}
Using the fact that the decay rate is RG invariant, i.e., that $dK/d\ln\mu=0$, the second factor may be written in analogy with (\ref{Dcircle}) as
\begin{equation}\label{KHdef}
   K_H(m_H^2)
   = \alpha_s^2(\mu^2)\,b^2(\mu^2)\,
    \frac{\pi}{2m_H^4}\,\mbox{Im}\Pi_{gg}(-m_H^2+i\epsilon,\mu^2)
   = \frac{1}{2\pi i} \oint\limits_{|s'|=m_H^2}\!\frac{ds'}{s'}\,
   D_H(-s') \,.
\end{equation}
The perturbative series of the function $D_H(Q^2)$ takes the form
\begin{equation}\label{DHexp}
   D_H(Q^2) = \alpha_s^2(Q^2) \left[ 1 + \sum_{n=1}^\infty d_n^H 
   \left( \frac{\alpha_s(Q^2)}{4\pi} \right)^n \right] ,
\end{equation}
where the expansion coefficients $d_n$ are pure numbers, independent of the renormalization scale. At NLO we have \cite{Baikov:2006ch}
\begin{equation}
   d_1^H = 73 - \frac{14}{3}\,n_f + \frac{2\beta_1}{\beta_0} 
   = 7\beta_0 - 4 + \frac{2\beta_1}{\beta_0} \,.
\end{equation}
In fixed-order perturbation theory, the perturbative series for the function $K_H(m_H^2)$ up to ${\cal O}(\alpha_s^4)$ takes the same form as (\ref{DHexp}) but with $\alpha_s(Q^2)$ replaced by $\alpha_s(m_H^2)$.  The corresponding expression in RG-improved perturbation theory is obtained by inserting the expansion (\ref{DHexp}) into (\ref{KHdef}) and using the contour integrals from Appendix~\ref{app2}. At NLO we find
\begin{equation}
   K_H(m_H^2) = \frac{\alpha_s^2(m_H^2)}{1+a^2}
   \left\{ 1 + \frac{1}{1+a^2}\,\frac{\alpha_s(m_H^2)}{4\pi} 
   \left[ d_1^H + \frac{\beta_1}{\beta_0}
   \left( (1-a^2)\,\frac{\arctan(a)}{a} \!-\! \ln(1+a^2) \!-\! 1 \right) 
   \right] \right\} \! ,
\end{equation}
where $a\equiv\beta_0\alpha_s(m_H^2)/4$. For $a=0.2$, we find numerically
\begin{equation}
   K_H(m_H^2) = 0.962\,\alpha_s^2(m_H^2)
   \left\{ 1 + \frac{\alpha_s(m_H^2)}{4\pi} \left[ 
   0.962\,d_1^H - 0.088\,\frac{\beta_1}{\beta_0} \right] \right\} ,
\end{equation}
corresponding to a reduction of the LO and NLO terms by 4\% and 8\%, respectively.

As a final comment, we emphasize again the different nature of the perturbative series for Higgs-boson production compared with that for Higgs-boson decay or the $e^+ e^-\to\mbox{hadrons}$ cross section. In the former case, $\pi^2$-enhanced perturbative corrections resulting from Sudakov double logarithms arise already at first order in perturbation theory and are due to the cusp anomalous dimension governing the RG evolution of the hard component of the hard-scattering kernel $C$. Terms proportional to $n_f$, which reflect the evolution of the QCD coupling constant, enter first at two-loop order. The scale dependence of the $\alpha_s^2(\mu^2)$ factor in the Born-level cross section is compensated by that of the gluon-gluon luminosity function. For the cases of Higgs-boson decay or the $e^+ e^-\to\mbox{hadrons}$ cross section, on the other hand, the perturbative series are free of $\pi^2$-enhanced contributions up to NNLO, while $n_f$-dependent corrections arise already at first non-trivial order in the expansion in $\alpha_s$. One may therefore argue that the resummation of terms of the form $\beta_0^n\alpha_s^{n+1}$ is more important in these cases than the resummation of $\pi^2$-enhanced contributions. In fact, the leading such effects can be absorbed by setting the renormalization scale in the running coupling according to the BLM prescription \cite{Brodsky:1982gc}. The leading ${\cal O}(\beta_0\alpha_s^2)$ correction in the series for the $e^+ e^-\to\mbox{hadrons}$ cross section is absorbed by using the scale $\mu_{\rm BLM}=e^{2\zeta_3-11/4}\,\sqrt{s}\approx 0.708\,\sqrt{s}$ in the running coupling constant. In the case of Higgs-boson decay one should use $\mu_{\rm BLM}=e^{-7/4}\,m_H\approx 0.174\,m_H$. The resummation of BLM terms to all orders and the associated renormalon ambiguities for the scalar correlator were analyzed in \cite{Broadhurst:2000yc}, where also the resummation of the $\pi^2$-enhanced terms was discussed.

\section{Conclusions} 
\label{sec:con}

We have presented a RG-improved prediction for the total Higgs-boson production cross section. Our result is based on a factorization theorem for the partonic cross section near threshold, which is obtained by considering a sequence of effective theories in which the contributions associated with higher scales are successively integrated out. In a first step the top-quark is removed and an effective $Hgg$ interaction is derived. Subsequently, hard and soft modes are integrated out and absorbed into Wilson coefficients $H$ and $S$ defined in SCET. This separation of the physics associated with different scales enables us to evaluate the different contributions at their optimal scale, where they have a well-behaved perturbative expansion. After this is done, the different elements are evolved to a common scale by solving the RG equations obeyed by the Wilson coefficients. 

The hard function $H$ is given by the square of the on-shell gluon form factor, and we observe that this function receives large perturbative corrections due to $\pi^2$ terms arising in the analytic continuation of this form factor to time-like momentum transfer. These large corrections can be avoided by evaluating the hard function in terms of an expansion in powers of the running coupling at time-like momentum transfer. While the choice $\mu_h^2=-m_H^2$ is unconventional, it is natural in the sense that it minimizes the logarithms in the hard function. The RG then sums up the logarithms that arise in the evolution between different scales. In our case, part of this evolution takes place in the complex $\mu^2$-plane and is driven by the cusp anomalous dimension, and the logarithms which are resummed are Sudakov double logarithms of $-1$. Our results for the resummed cross sections are stable under variation of the matching scales. Compared to the fixed-order expansion, the resummed perturbative series exhibits smaller theoretical uncertainties and better convergence. This improvement is mainly due to the resummation of the $\pi^2$-enhanced terms, while soft-gluon resummation alone has a small impact. We find that the effects of RG improvement are significant even at NNLO. For example, for $m_H=120$\,GeV the resummed NNLO cross sections at the Tevatron and the LHC exceed the fixed-order predictions by about 13\% and 8\%, respectively. We perform a detailed analytical and numerical comparison with the traditional moment-space approach to resummation. While formally equivalent, the two methods differ by terms that are power suppressed in the threshold region. We analyze the difference in detail and show that it explains the somewhat larger threshold resummation effects found using the traditional method.

We also discussed RG improvement for a number of other time-like processes, for which Sudakov logarithms are absent and our method reduces to contour-improved perturbation theory. The effects of choosing a time-like renormalization scale are modest for the cases of the total $e^+ e^-\to\mbox{hadrons}$ cross section and for the hadronic decay rate of the Higgs boson. Because these inclusive processes do not contain double-logarithmic corrections, the $\pi^2$-enhanced contributions arise solely from the running of $\alpha_s$ and start at NNLO in fixed-order perturbation theory. These effects are significant for hadronic $\tau$ decays, because the characteristic scale is rather low. For Higgs-boson decay, on the other hand, we have argued that BLM-type corrections of ${\cal O}(\beta_0^n\alpha_s^{n+1})$ may be more important phenomenologically.

It will be interesting to extend the methods developed here to other hard-scattering processes, such as jet production or heavy-quark production at hadron colliders. In general, processes containing collinear partons in both the initial and final states of the collision are characterized by an interplay of both time-like and space-like momentum transfers, so that the scale separation in the effective theory becomes more complicated. We hope to return to such issues in future work.

\newpage

\vspace{0.2cm}
{\em Acknowledgments:\/}
We are grateful to Babis Anastasiou, Martin Beneke, Kirill Melnikov, Frank Petriello, Giulia Zanderighi, and Jos\'e Zurita for useful discussions. We thank Robert Schabinger for pointing out a misprint in equation~(A.3) in a former version of this paper, which we have now corrected. M.N.\ thanks the Fermilab Theory Group and T.B.\ thanks the ITP at the University of Z\"urich for hospitality and support during the final stages of this work. The research of T.B.\ was supported by the U.S.\ Department of Energy under Grant DE-AC02-76CH03000. Fermilab is operated by the Fermi Research Alliance under contract with the Department of Energy.

\begin{appendix}

\section{Anomalous dimensions}
\label{app1}
\renewcommand{\theequation}{A\arabic{equation}}
\setcounter{equation}{0}

We write the perturbative expansions of the anomalous dimensions and $\beta$-function in the form
\begin{equation}
   \gamma(\alpha_s)
   = \sum_{n=0}^\infty\,\gamma_n
    \left( \frac{\alpha_s}{4\pi} \right)^{n+1} , \qquad
   \beta(\alpha_s) 
   = -2\alpha_s \sum_{n=0}^\infty\,\beta_n 
    \left( \frac{\alpha_s}{4\pi} \right)^{n+1} .
\end{equation}
The cusp anomalous dimension in the adjoint representation is given (at least up to three-loop order) by $C_A/C_F$ times that in the fundamental representation. The first four expansion coefficients of the cusp anomalous dimension and the $\beta$-function can be found, e.g., in Appendix~B of \cite{Becher:2007ty}, where also the explicit expressions for the evolution functions $S$ and $a_\Gamma$ in (\ref{Saints}) valid at NNLO are summarized.

The first three expansion coefficients of the anomalous dimension $\gamma^S$ entering the evolution equation (\ref{CSevol}) of the hard matching coefficient $C_S$ read \cite{Idilbi:2005ni,Idilbi:2005er}
\begin{equation}
\begin{aligned}
  \gamma^S_0 &= 0 \,, \\
  \gamma^S_1 
  &= C_A^2 \left( -\frac{160}{27} + \frac{11\pi^2}{9}
   + 4\zeta_3 \right) 
   + C_A T_F n_f \left( -\frac{208}{27} - \frac{4\pi^2}{9} \right) 
   - 8 C_F T_F n_f \,, \\
  \gamma^S_2 
  &= C_A^3 \left[ \frac{37045}{729} + \frac{6109\pi^2}{243}
   - \frac{319\pi^4}{135} 
   + \left( \frac{244}{3} - \frac{40\pi^2}{9} \right) \zeta_3 
   - 32\zeta_5 \right] \\
  &\quad\mbox{}+ C_A^2 T_F n_f \left( -\frac{167800}{729}
   - \frac{2396\pi^2}{243} + \frac{164\pi^4}{135} 
   + \frac{1424}{27}\,\zeta_3 \right) \\
  &\quad\mbox{}+ C_A C_F T_F n_f \left( \frac{1178}{27}
   - \frac{4\pi^2}{3} - \frac{16\pi^4}{45} 
   - \frac{608}{9}\,\zeta_3 \right) + 8 C_F^2 T_F n_f \\  
  &\quad\mbox{}+ C_A T_F^2 n_f^2 \left( \frac{24520}{729}
   + \frac{80\pi^2}{81} - \frac{448}{27}\,\zeta_3 \right) 
   + \frac{176}{9} C_F T_F^2 n_f^2 \,.
\end{aligned}
\end{equation}
The first three coefficients of the anomalous dimension $\gamma^B$, which equals one half of the coefficient of the $\delta(1-x)$ term in the Altarelli-Parisi splitting function $P_{g\leftarrow g}(x)$, are \cite{Vogt:2004mw}
\begin{equation}
\begin{aligned}
  \gamma^B_0 
  &= \frac{11}{3}\,C_A - \frac{4}{3}\,T_F n_f
   = \beta_0 \,, \\
  \gamma^B_1 
  &= 4 C_A^2 \left( \frac{8}{3} + 3\zeta_3 \right) 
   - \frac{16}{3}\,C_A T_F n_f - 4 C_F T_F n_f \,, \\
  \gamma^B_2 
  &= C_A^3 \left[ \frac{79}{2} + \frac{4\pi^2}{9} + \frac{11\pi^4}{54} 
   + \left( \frac{536}{3} - \frac{8\pi^2}{3} \right) \zeta_3 
   - 80\zeta_5 \right] \\
  &\quad\mbox{}- C_A^2 T_F n_f \left( \frac{233}{9} + \frac{8\pi^2}{9}
   + \frac{2\pi^4}{27} + \frac{160}{3}\,\zeta_3 \right) \\
  &\quad\mbox{}- \frac{241}{9}\,C_A C_F T_F n_f + 2 C_F^2 T_F n_f
   + \frac{58}{9}\,C_A T_F^2 n_f^2 + \frac{44}{9}\,C_F T_F^2 n_f^2 \,.
\end{aligned}
\end{equation}

\section{Contour integrals}
\label{app2}
\renewcommand{\theequation}{B\arabic{equation}}
\setcounter{equation}{0}

Here we evaluate the contour integrals
\begin{equation}
   I_n(a) 
   = \frac{1}{2\pi i} \oint\limits_{|s'|=s}\!\frac{ds'}{s'}
    \left( \frac{\alpha_s(-s')}{\alpha_s(s)} \right)^n
   = \frac{1}{2\pi} \int_{-\pi}^\pi\!d\varphi
    \left( \frac{\alpha_s(e^{i\varphi} s)}{\alpha_s(s)} \right)^n ,
\end{equation}
where $a\equiv\beta\alpha_s(s)/4$. Using relation (\ref{ascomplex}) for the running coupling in the complex momentum plane, we find at NLO
\begin{equation}
   I_n(a) = F(\eta)\Big|_{\eta\to n-1} 
   + \frac{n\beta_1}{\beta_0}\,\frac{\alpha_s(s)}{4\pi}\,
   F'(\eta)\Big|_{\eta\to n} + \dots \,,
\end{equation}
where
\begin{equation}
   F(\eta) = \frac{i}{2a\eta} 
   \left[ \left( 1+ia \right)^{-\eta} - \left( 1-ia \right)^{-\eta} 
   \right] .
\end{equation}
Explicitly, we obtain for the first three integrals
\begin{equation}\label{Ins}
\begin{aligned}
   I_1(a) &= \frac{\arctan(a)}{a}
    + \frac{\beta_1}{\beta_0}\,\frac{\alpha_s(s)}{4\pi}\,
    \frac{1}{1+a^2} \left[ \frac{\arctan(a)}{a} - \frac12\ln(1+a^2)
    - 1 \right] , \\
   I_2(a) &= \frac{1}{1+a^2} 
    + \frac{\beta_1}{\beta_0}\,\frac{\alpha_s(s)}{4\pi}\,
    \frac{1}{\left(1+a^2\right)^2} \left[ 
    (1-a^2)\,\frac{\arctan(a)}{a} - \ln(1+a^2) - 1 \right] , \\
   I_3(a) &= \frac{1}{\left(1+a^2\right)^2} 
    + \frac{\beta_1}{\beta_0}\,\frac{\alpha_s(s)}{4\pi}\,
    \frac{1}{\left(1+a^2\right)^3} \left[ 
    (1-3a^2)\,\frac{\arctan(a)}{a} - \frac{3-a^2}{2} \ln(1+a^2) - 1
    + \frac{a^2}{3} \right] .
\end{aligned}
\end{equation}

\end{appendix}

\end{document}